\documentclass[11pt]{article}
\usepackage{latexsym, amssymb, amsmath, amsthm, citesort}
\usepackage{graphicx}
\usepackage{subfigure}
\usepackage{epsfig}
\usepackage{enumerate}
\usepackage{epstopdf}

\theoremstyle{plain}

\newcommand{\R}{\mathbf{R}}

\newcommand{\e}{\mathrm{e}}
\renewcommand{\i}{\mathrm{i}}
\def\E{{\hbox{\bf E}}}

\renewcommand{\>}{\rangle}

\numberwithin{equation}{section}
\title{Detecting Highly Oscillatory Signals by Chirplet Path Pursuit}

\author{Emmanuel J. Cand\`es$^{\dagger}$, Philip R. Charlton$^{\sharp}$
  and Hannes Helgason$^{\dagger}$\\
  \vspace{-.3cm}\\
  $\dagger$ Applied and Computational Mathematics, Caltech, Pasadena, CA 91125, USA\\
  \vspace{-.4cm}\\
  $\sharp$ School of Science and Technology, Charles Sturt University,\\
  Wagga Wagga, NSW 2678, Australia}

\date{March 2006} 

\begin{document}
\maketitle

\begin{abstract}
  This paper considers the problem of detecting nonstationary
  phenomena, and chirps in particular, from very noisy data. Chirps
  are waveforms of the very general form $A(t) \exp(\i\lambda \,
  \varphi(t))$, where $\lambda$ is a (large) base frequency, the phase
  $\varphi(t)$ is time-varying and the amplitude $A(t)$ is slowly
  varying. Given a set of noisy measurements, we would like to test
  whether there is signal or whether the data is just noise. One
  particular application of note in conjunction with this problem is
  the detection of gravitational waves predicted by Einstein's Theory
  of General Relativity.
  
  We introduce detection strategies which are very sensitive and more
  flexible than existing feature detectors. The idea is to use
  structured algorithms which exploit information in the so-called
  chirplet graph to chain chirplets together adaptively as to form
  chirps with polygonal instantaneous frequency. We then search for
  the path in the graph which provides the best trade-off between
  complexity and goodness of fit.  Underlying our methodology is the
  idea that while the signal may be extremely weak so that none of the
  individual empirical coefficients is statistically significant, one
  can still reliably detect by combining several coefficients into a
  coherent chain. This strategy is general and may be applied in many
  other detection problems.  We complement our study with numerical
  experiments showing that our algorithms are so sensitive that they
  seem to detect signals whenever their strength makes them
  detectable.
\end{abstract}


{\bf Acknowledgments.} E.~C. was partially supported by National
Science Foundation grants DMS 01-40698 (FRG) and ITR ACI-0204932.
P.~C. was partially supported by NSF grant PHY-0107417.
Many thanks to David Donoho, Houman Ohwadi, Justin Romberg and Chiara
Sabatti for fruitful conversations. We would also like to thank Ery
Arias-Castro for references. The results in this paper were first
presented at ``Regularization in Statistics,'' Banff, Canada,
September 2003 and at ``Multiscale Geometric Analysis in High
Dimensions,'' UCLA, Los Angeles, California, November 2004
\cite{IPAM}.

\pagebreak

\section{Introduction}
\label{sec:introduction}

This paper considers the problem of detecting one-dimensional signals
from noisy measurements.  Suppose we have noisy sampled data
\begin{equation}
\label{eq:model}
y_i = \alpha \, S_i + z_i, \qquad i = 1, \ldots, N, 
\end{equation}
where the unknown vector $(S_i)$ are sampled values $S_i = S(t_i)$ of
a signal of interest $S(t), t \in [0,1]$, and $(z_i)$ is a zero-mean
stochastic vector, not necessarily i.i.d. but with a known
distribution. Based on the observations $(y_i)$, one would like to
decide whether or not a signal is hiding in the noise. Formally, we
would like to test the null hypothesis 
\[
H_0 : \alpha = 0 \mbox{ (noise only)}
\]
against the alternative
\[
H_1 : \alpha \neq 0 \mbox{ (signal is buried in noise)}.
\]
We emphasize here is that the signal $S$ is completely unknown and may
not depend upon a small number of parameters. This situation is
commonly referred as {\em nonparametric testing} as opposed to the
classical parametric setup where it is assumed that the set of
candidate signals belong to a known parametric class. 

\subsection{Chirps}

In this paper, we will focus our attention on the detection of
frequency modulated signals also named `chirps.' Roughly speaking, a
chirp is a signal of the form
\begin{equation}
  \label{eq:chirp}
  S(t) = A(t) \, \cos(\lambda \varphi(t)), 
\end{equation}
where the amplitude $A$ and the phase $\varphi$ are smoothly varying
functions of time, and where the oscillation degree $\lambda$ is
large. It follows from the definition that chirps are highly
oscillatory signals with a frequency content $\omega(t)$ 
also rapidly changing over time; although this is an ill-defined
concept, researchers like to talk about the `instantaneous frequency'
of a chirp simply defined as the derivative of the phase function
\begin{equation}
  \label{eq:instantaneous}
  \omega(t) = \lambda \varphi'(t). 
\end{equation}
(This definition can be justified in the case where $\lambda
|\varphi'(t)|^2/|\varphi''(t)| \gg 1$ and we skip the details
\cite{MeyerLewis}.)  Hence, for large values of $\lambda$, the
frequency content of a chirping signal is also rapidly changing with
time.

Chirps arise in a number of important scientific disciplines,
including the analysis of Echolocation in Bats \cite{Bats1,Bats2}
and other mammals \cite{Whales}, the study of atmospheric whistlers
\cite{Whistlers1}, and very recently, in efforts to detect
gravitational waves
\cite{thorne:300yearsofgravitation,anderson:tfgwdetection}.
For example, a particular species of bats ({\em Eptesicus fuscus}) uses a
remarkable sonar system for navigation in which some specific chirps
are emitted. Because chirps are ubiquitous in nature, strategies for
their detection are bound to be of great practical interest.  We
briefly mention two applications:
\begin{enumerate}
\item {\em Remote sensing.} Suppose that we have one or several
  objects moving in a cluttered background. In anti submarine warfare,
  for example, one would like to detect the presence of submarines
  from noisy acoustic data. Because different engines have different
  time-frequency characteristics, we would expect the signal at the
  sensor to behave like a chirp.  In a more peaceful underwater
  setting, whales are known to emit chirping sounds \cite{Whales} and
  their detection would help locating and/or tracking these
  mammals. If we could also estimate some basic characteristics of the
  chirp, then one could also discern between different types of
  submarines, or different species of whales.

  A closely related application is active ranging where one detects
  the location and velocity of objects by sending an electromagnetic
  wave and recording the echo. Because objects are moving, the Doppler
  effect implies that the recorded signal is chirp-like.

\item {\em Detection of gravitational waves.}  Whereas the existence
  of gravitational waves was predicted long ago by the theory of
  general relativity, they have not been confirmed experimentally
  \cite{thorne:300yearsofgravitation}. There are massive ongoing
  efforts aimed at detecting gravitational waves. Detecting
  gravitational waves on earth is very challenging because of the
  extremely tiny effects they induce on physical systems so that the
  signal is expected to be buried in a ¡Æsea of
  noise.¡Ç Interestingly, many gravitational waves are well-modeled by
  time-frequency modulated signals
  \cite{anderson:tfgwdetection,allen}.
\end{enumerate}
 
\subsection{Gravitational waves}
\label{sec:gw}

In this short section, we briefly expand on the problem of detecting
gravitational waves as this really is the main applicative focus of
this line work, see the companion paper \cite{GWs}. Predicted by
Einstein's General Theory of Relativity, gravitational waves are
disturbances in the curvature of spacetime caused by the motions of
matter. Expressed in a different way, they are oscillations
propagating at the speed of light in the fabric of spacetime.
A strong source of gravitational waves is a system made up of two
massive objects closely and rapidly orbiting each other, e.g. two
black holes, two neutron stars, etc. As they are orbiting, general
relativity predicts that the system loses energy in the form of
gravitational radiation, causing the objects to gradually spiral in
towards each other, eventually resulting in a violent merger.

When a gravitational passes, it alternatively stretches and shrinks
distances. The Laser Interferometric Gravitational-wave Observatory (LIGO)
\cite{LIGO} is a sophisticated apparatus which uses interferometry to
detect such variations (VIRGO is the European equivalent). The
difficulty is that even strong sources will induce variations in an
incredibly small scale: the variation is proportional to the length
\[
\Delta L = h L,
\]
where the factor $h$ is about $10^{-21}$. In LIGO, $L$ is the distance
between two test masses and is about 4 kilometers so that one would
need to detect a displacement of about $4 \times 10^{-18}$
meters. This is a formidable challenge as this distance is about 1,000
times smaller than the nucleus of an atom! Because of many other
sources of variations, the measurements of the displacements are
expected to be extremely noisy.

Gravitational wave astronomy could offer a new window to the universe
and expand our knowledge of the cosmos dramatically.  Aside from
demonstrating the existence of black holes and perhaps providing a
wealth of data on supernovae and neutron stars, gravitational wave
observations could revolutionize our view of the universe and unveil
phenomena never considered before. We quote from Kip Throne, one of
the leading LIGO scientists \cite{LIGO}: ``Gravitational-wave detectors
will soon bring us observational maps of black holes colliding [...]
and black holes thereby will become the objects of detailed
scrutiny. What will that scrutiny teach us? There will be surprises.''

To see the relevance of our problem to gravitational wave detection,
one can use Einstein's equations to predict the shape of a wave. In
the case of the coalescence of a binary system, the gravitational wave
strain $S(t)$ (the relative displacement as a function of time) is
approximately of the form
\[
S(t) = A \cdot (t_c - t)^{-1/4} \, \cos(\omega \, (t - t_c)^{5/8} +
\phi), 
\]
where $A$ is a constant, $\omega$ is large and $t_c$ is the time of
coalescence (merger); this approximation is only valid for $t < t_c$
and we do not know the shape of $S$ after the merger. Clearly, $S(t)$
is an example of a chirp. It is possible to push the calculations and
add corrective terms to both the phase and amplitude
\cite{anderson:tfgwdetection,allen}. 

\subsection{The challenges of chirp detection}

While the literature on nonparametric estimation is fairly developed,
that of nonparametric is perhaps more recent, see
\cite{Ingster:NonparamDetectionBook} and references therein. A typical
assumption in nonparametric detection is to assume that the signal $S$
is smooth or does not vary too rapidly. For example, one could define
a class of candidate signals by bounding the modulus of smoothness of
$S$, or by imposing that $S$ lies in a ball taken from a smoothness
class, e.g. a Sobolev or Besov type class. In this paper and
specializing, \eqref{eq:chirp}, one might be interested in knowing
whether or not a signal such as $S(t) = \cos (\pi N t^2/2)$ for $t \in
[0,1]$ is hiding in the data. If one collects samples at the
equispaced time points $t_i = (i-1)/N$, then one can see that the
signal changes sign at nearly the sampling rate (note that the
frequency content is also changing at the sampling rate since the
`instantaneous frequency' increases linearly from $\omega = 0$ to
$\omega = \pi N$). With more generality, one could introduce a
meaningful class of chirps of the form $A(t) \, \cos(N \varphi(t))$
by imposing the condition that $A$ and $\varphi$ have bounded higher order
derivatives.  The behavior of signals taken from such classes is very
different than what is traditionally assumed in the literature. This
the reason why the methodology we are about to introduce is also
radically different.


Many of the methods envisioned for detecting gravitational waves heavily
rely on very precise knowledge of the signal waveforms. The idea is to
approximate the family of signals with a finite collection $\{S_\theta
: \theta \in \Theta\}$ and perform a sort of generalized likelihood
ratio test (GLRT) also known as the method of matched filters
\cite{owen:matchedfiltering}. For example, in additive Gaussian white
noise, one would compute $Z^* = \max_{\theta \in \Theta} \<y,
S_\theta\>/\|S_\theta\|$ and compare the value of the statistic with a
fixed threshold. This methodology suffers from some severe problems
such as prohibitive computational costs and lack or robustness
resulting in poor detection methods where the waveforms are not well
known. More to the point, this assumes that the unknown signal may be
reduced to a finite set of templates so that the problem essentially
reduces to a parametric test, which we are not willing to assume in
this paper. Instead, and keeping Thorne's comment in mind, one would
like a robust and flexible methodology able to detect poorly modeled
or even totally unknown but coherent signals.

\subsection{Current detection strategies}

In the last decade or so, various time-frequency methods have been
proposed to overcome the problems with matched filters. For example,
in \cite{Torresani:ChirpDetection}, the proposal is to look for ridges
in the time-scale plane. One computes the continuous time wavelet
transform $W(a,b)$ where $a > 0$ is scale and $b$ is time and search
for a curve $\rho(a)$ along which the sum of $\int |W(a,\rho(a))|^2
da/a$ is maximum. Because one cannot search the whole space of curves,
the method is restricted to parametric `power-law chirps' of the form
$S(t) = (t_0 - t)_+^\alpha \cos(2\pi F_\beta (t_0-t)^{\beta+1})$ where
$\alpha$ and $\beta$ are unknown and $F_\beta$ is some constant. This
is again a parametric problem. While the approach is more robust than
the method of matched filters, it is also less sensitive. Among other
issues, one problem is that the wavelet transform does not localize
energy as much as one would want. In a similar fashion,
\cite{ChassandreFlandrin} proposes to search for ridges in the
time-frequency plane. Again, the methodology is designed for power-law
chirps. The idea is to use appropriate time-frequency distributions
such as the Wigner-Ville distribution (WVD) to localize the unknown
signal as much as possible in the time-frequency plane. Here one
needs, different time-frequency distributions for different chirp
parameters $\alpha$ and $\beta$. For example the WVD is ideal for
linear chirps but is ill-adapted to hyperbolic chirps, say, and one
has to use another distribution. In practice, one also needs to deal with the
undesirable interference properties of the WVD; to fix the
interference methods requires ad-hoc methods which, in turn, have
consequences on the sensitivity of the detector
\cite{ChassandreFlandrin}. To summarize, while these methods may be
more robust in parametric setups, they are also far less powerful.

As far as nonparametrics is concerned, \cite{anderson:tfgwdetection}
introduces a strategy where as before, the idea is to transform the
data via the WVD giving the time frequency distribution
$\rho(t,\omega)$, and then search for ridges (which is a problem
similar to that of finding edges from very noisy image data). A
decision is made whether or not each point in the time-frequency plane
is a `ridge point,' and the value of the statistical test depends on
the length of the longest ridges.  Again, because the WVD of a clean
signal can take nonzero values in regions of the time-frequency plane
having nothing to do with the spectral properties of the signal, one
has to use a smoothing kernel to average the interference patterns,
which simultaneously smears out the true ridges, thereby causing a
substantial loss of resolution. Another issue with this approach is
that while the signal may be detectable, it may not be locally
detectable. This means that a huge majority of true ridge points may
lie in the bulk of the data distribution, and not pass the threshold. 

\subsection{This paper}

In this paper, we propose a detection strategy, which is very
different than those currently employed in the literature. As
explained earlier, we cannot hope to generate a family of chirps that
would provide large correlations with the unknown signal; indeed, for
a signal of size $N$, one would need to generate exponentially many
waveforms which is unrealistic.  But it is certainly possible to
generate a family of templates which provide good {\em local} correlations,
e.g. over shorter time intervals.  We then build the so-called family
of {\em chirplets} and our idea is to use a structured algorithm which
exploit information in the family to chain chirplets together
adaptively as to form a signal which is physically meaningful, and
whose correlation with the data is largest. The basic strategy is as
follows:
\begin{itemize}
\item We compute empirical coefficients by correlating the data with a
  family of templates which is rich enough to provide good local
  correlations with the unknown signal.

\item We then exploit these empirical correlations and chain our
  coefficients in a meaningful way; i.e. so that the chain of
  `templates' approximates a possible candidate signal.

\item A possible detection strategy might then compare the sum of
  local correlations along the chain with a fixed threshold.
\end{itemize}
Of course, one would need to specify which templates one would want to
use, which chaining rules one would want to allow, and how one should
use this information to decide whether any signal is present or
not. This is the subject of the remainder of this paper. But for now,
observe that this is a general strategy. Suppose for simplicity that
in the data model \eqref{eq:model}, the error terms are
i.i.d. $N(0,\sigma^2)$, and denote by $(f_v)_{v \in V}$ our class of
templates. One can think of our strategy as computing the 
individual $Z$-scores
\[
Z_v = \<y,f_v\>/\|f_v\|, \quad Z_v \sim N(\mu_v, \sigma^2), 
\]
where $\mu_v = \alpha \<S, f_v\>/\|f_v\|$, and then searching for a
chain of templates such that the sum of the $Z$-scores is large. The
key point here is that while the signal-to-noise ratio may be so low
so that none of the individual $Z$ scores achieves statistical
significance, $\mu_v \ll \sigma$ (i.e. the signal is there but not
locally detectable), their sums along carefully selected paths would
be judged statistically significant so that one could detect reliably.

We use the word `path' because we have a graph structure in mind.  In
effect, we build a graph $G = (V, E)$ where the nodes $V$ correspond
to our set of templates $(f_v)_{v \in V}$ and the edges are
connectivities between templates. The edges are chosen so that any
path in the graph approximates a meaningful signal. The idea is to
find a path which provides the best trade-off between correlation and
complexity, where our measure of complexity is the number of templates
used to fit the data. 

Our methods are adaptive in the sense that they do not require any
precise information about the phase and amplitude and yet, they are
efficient at detecting a wide family of nonparametric
alternatives. Our methods are also versatile and can accommodate many
different types of noise distributions. For example, we will examine
the case where the noise is Gaussian stationary with a known spectrum
since this is the assumed noise distribution in gravitational wave detectors
such as LIGO. Last but not least, our methods have remarkably low
computational complexity, which is crucial for their effective
deployment in applications.



\subsection{Inspiration}

Our methods are inspired by \cite{Donoho:Beamlets} where it was
suggested that one could chain together beamlets to detect curves from
noisy data. To the best of our knowledge, the idea of finding paths in
a locally connected network to isolate salient features goes back to
Sha'ashua and Ullman \cite{Ullman}. Closely related to this line of
work is the more recent work \cite{Arias:FilamentDetection} where a
graph structure is used to detect filamentary structures.  Having said
that, the literature on the use of graphical models in signal
detection is short. The detection strategies in this paper are
different than those presented in the aforementioned references and
are, therefore, adding to the developing literature. They can be
applied to the problem of detecting chirps and gravitational waves in
Astronomy but it is clear that they are also very general and can be
tailored to address a variety of other statistical problems as well. 

\section{Multiscale Chirplets and the Chirplet Graph}
\label{sec:chirpletgraph}

\subsection{Multiscale chirplets}

\begin{figure}
  \begin{center}
         \includegraphics[height=2.5in]{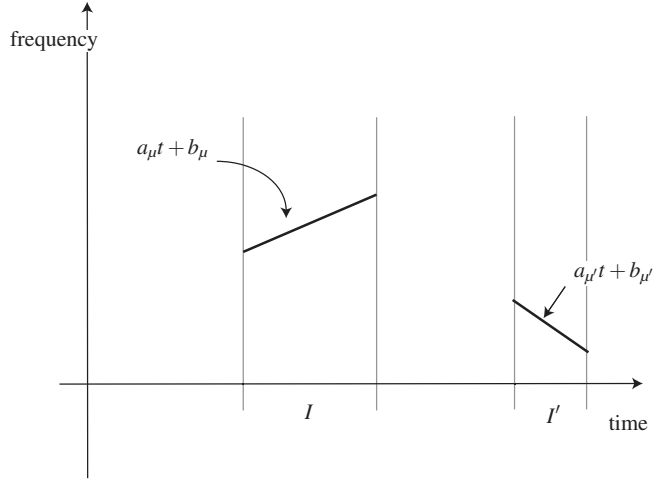}
  \end{center}
  \caption{\small Diagrammatic representation of two chirplets in the
    time-frequency plane.}
  \label{fig:chirplet}
\end{figure}

This section introduces a family of multiscale chirplets which provide
good local approximations of chirps under study. We assume we work in
the time interval $[0,1]$ (and that our measurements are evenly
sampled), and for each $j \ge 0$, we let $I$ denote the dyadic
interval $I = [k 2^{-j}, (k+1) 2^{-j}]$, where $k = 0, 1, \ldots, 2^j
- 1$. We then define the multiscale chirplet dictionary as the family
of functions defined by
\begin{equation}
  \label{eq:chirplet}
  f_{I,\mu}(t) := |I|^{-1/2} \, e^{\i (a_\mu t^2/2 + b_{\mu} t)} \, 1_I(t),
\end{equation}
where $(a_\mu, b_\mu) \in {\cal M}_j$ is a discrete collection of offset and
slope parameters which may depend on scale and on prior information
about the objects of interest. We note that thanks to the
normalization factor, chirplets are unit-normed, $\|f_{I,\mu}\|_{L_2}
= 1$.  This system is appealing because a time-frequency portrait of
its elements (say, by Wigner-Ville Distribution) reveals a system of
elements of all possible durations, locations, average frequencies,
and, most importantly, chirprates which are linear changes of
instantaneous frequency during the interval of operation.  Indeed, one
can think of the `instantaneous frequency' of a chirplet as being
linear and equal to $a_{\mu} t + b_{\mu}$ so that in a diagrammatic
sense, a chirplet is a segment in the time-frequency plane, see Figure
\ref{fig:chirplet}.

Chirp atoms were introduced to deal with the nonstationary behavior of
the instantaneous frequency of some signals.  The terminology
`chirplet' is borrowed from the work of Mann and Haykin
\cite{MannHaykin} (see also \cite{BaraniukJones}) who have proposed
the so-called chirplet transform of a signal: starting from the
Gaussian multiparameter collection of linear chirps
\begin{equation}
 \label{eq:gaussianchirplet}
g_{\lambda}(t) = g((t - b)/a) e^{\i (\omega t + \delta t^2)}, \quad 
\lambda = (a, b, \omega, \delta), 
\end{equation}
with $g$ a Gaussian window and $a > 0$, and $b, \omega, \delta \in
\R$, they define the chirplet transform of a signal $f$ as being the
collection of inner products $c_\lambda = \<f , g_\lambda\>$. The
resemblance between \eqref{eq:chirplet} and
\eqref{eq:gaussianchirplet} is self-evident---hence the
terminology. What is new here is the notion of chirplet graph, which
we will introduce below.


\subsection{Discretization}
\label{sec:discretization}

We give a possible discretization for an evenly sampled signal of size
$N = 2^J$. For each dyadic interval $I = [k2^{-j}, (k+1) 2^{-j}]$, we
mark out two vertical lines in the $[0,1] \times [-\pi, \pi]^2$ plane
at the endpoints of $I$, $t_I = k 2^{-j}$ and $t'_I = (k+1)2^{-j}$,
and place ticks along the vertical lines at spacing $2\pi/N$.  We then
create a dictionary of `chirplet lines' connecting tick marks. For the
simplicity of the exposition, suppose that the phase of the unknown
chirp is such that we only have to create such lines with a slope---in
absolute value---less or equal to $2 \pi$. (This is suitable whenever
$\lambda |\varphi''(t)| \le 2\pi \, N$.) A simple count shows that the
number of slopes is of size about $2 N \cdot 2^{-j}$ so that the
number $N_j$ of chirplets per dyadic interval obeys
\[
N_j = \# \, \text{offsets } \times \# \, \text{slopes} \approx N
\times 2 N/2^j.
\]  
In other words, there are about  
\[
2^j \, N_j \asymp 2 N^2 
\]
chirplets at scale $2^{-j}$. Thus, if we consider all scales $j = 0,
\ldots, \log_2(N) - 1$, we see that the size of this special chirplet
dictionary is about
\[
2 N^2 \log_2 N. 
\]

Of course, for evenly sampled signals, discrete chirplets are the
sampled waveforms $f_{I,\mu}[t] = f_{I,\mu}((t-1)/N)/\sqrt{N}$, with
$t = 1, \ldots, N$. With the discrete chirplet dictionary in hand, we
define the {\em chirplet analysis} or {\em transform} of a signal of
length $N$ as the collection of inner products with all the elements
in the dictionary.  We call these inner products chirplet
coefficients. It is clear that one can use the FFT to compute the
chirplet coefficient table. For example, with the above
discretization, it is possible to compute all the coefficients against
chirplets `living' in the fixed interval $[k 2^{-j}, (k+1)2^{-j})$ in
$O(N_j \log (N/2^j))$ flops so that the computational complexity of
the chirplet transform is $O(N^2 \log^2 N)$.  There are many other
possible discretizations and the experienced reader will also notice
that for regular discretizations, the complexity will scale as $O(M_N
\log N)$, where here and below $M_N$ is the number of chirplets in the
dictionary. In summary, the computational cost is at most of the order
$O(\log N)$ per chirplet coefficient.

\subsection{The chirplet graph}
\label{sec:chirpgraph}

\begin{figure}
  \begin{center}
    \begin{tabular}{ccc}
      \includegraphics[height=2.6in]{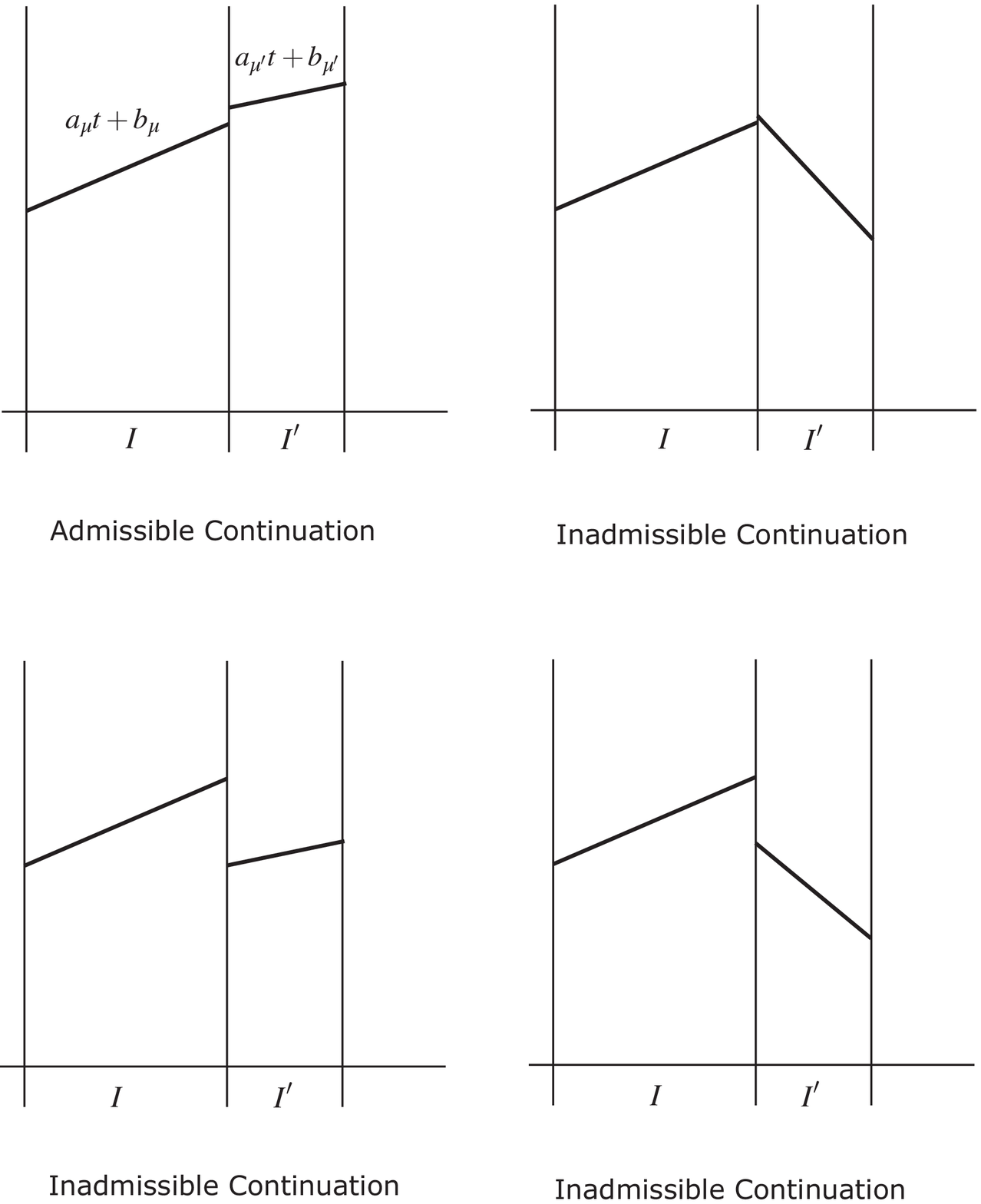} & & 
      \includegraphics[height=2.6in]{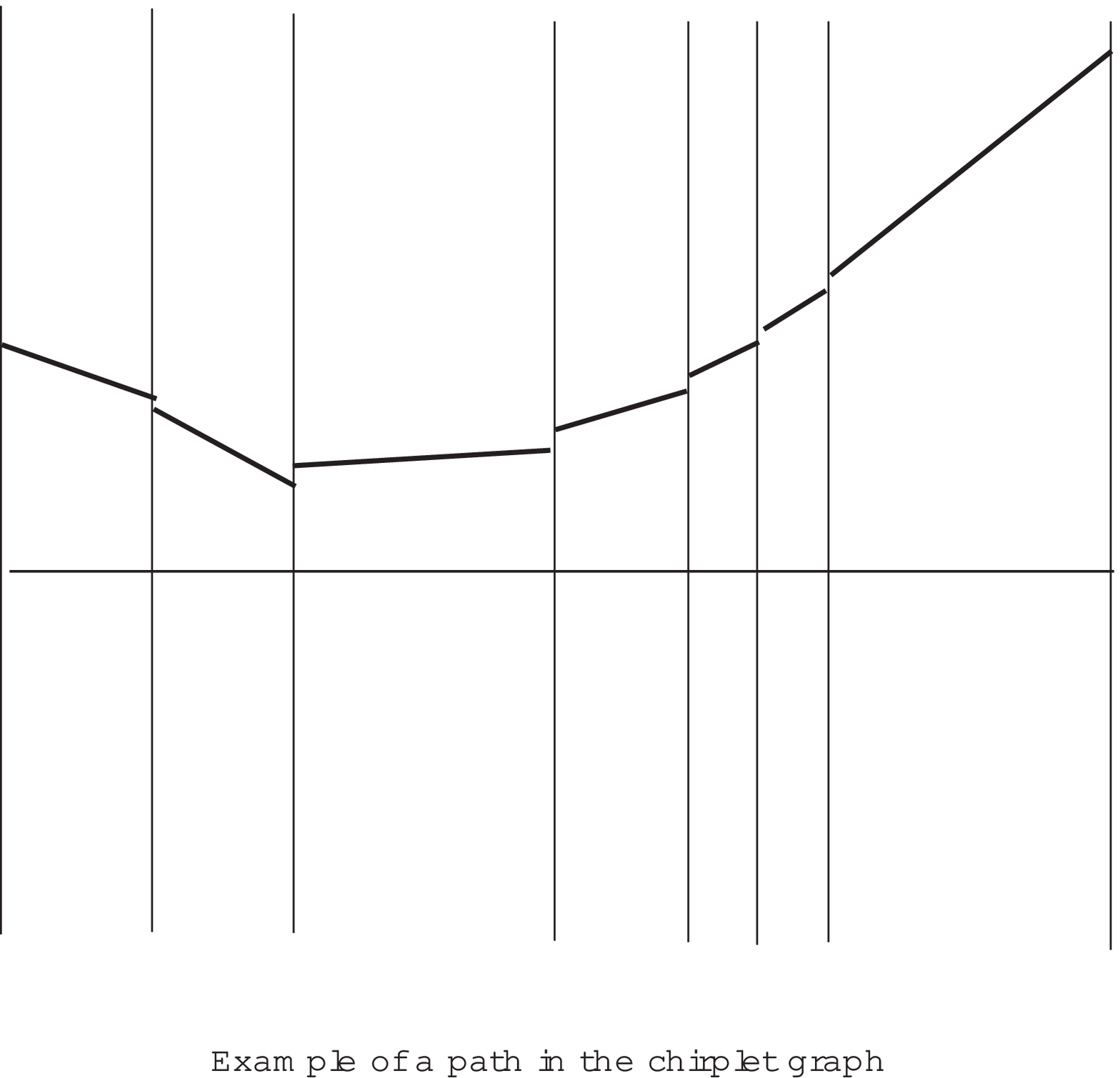}\\ (a) & & (b)
    \end{tabular}
  \end{center}
  \caption{\small (a) Example of connectivities in the chirplet
    graph. Chirplet may not be connected when the difference in
    offsets or slopes (or both) is large. (b) Diagrammatic
    representation of the instantaneous frequency along a path in the
    chirplet graph.}
  \label{fig:graph}
\end{figure}

The main mathematical architecture of this paper is the {\em chirplet
  graph} $G = (V, E)$ where $V$ is the set of nodes and $E$ the set of
edges.  Each node in the graph is a chirplet index $v = (I,\mu)$.
Throughout the paper, vertices corresponding to chirplets starting at
time $t=0$ are said to be start-vertices, and vertices corresponding
to chirplets ending at time $t=1$ are said to be end-vertices. The
edges between vertices are selected to impose a certain regularity
about the instantaneous frequency, see Figure \ref{fig:graph}.
\begin{enumerate}
\item First, a natural constriction is that two chirplets can only be
  connected if they have adjacent supports in time. 
\item Second, two chirplets are connected if the frequency offset at
  the juncture is small {\em and} if the difference in their slopes is
  not too large as well. 
\end{enumerate}
The idea is to model a chirp with instantaneous frequency $\lambda
\varphi'(t)$ as a sequence of connected line segments. Imposing
constraints on the connectivities is akin to imposing constraints on
the first derivative of the instantaneous frequency.  With these
definitions, a chirplet path is simply a set of connected vertices,
starting from a start-vertex and ending at an end-vertex.


\section{Detection Statistics}
\label{sec:teststatistic}

\newcommand{\cF}{{\cal F}}

We now describe the complete algorithm for searching chirps through
the data.  To explain our methodology, it might be best first to
focus on the case of additive Gaussian white noise
\[
y_i = \alpha S_i + z_ i, \quad i = 1, \ldots, N, \quad z_i \text{
  i.i.d. } N(0,\sigma^2).
\]
We wish to test $H_0 : \alpha = 0$ against $H_1 : \alpha \neq 0$.  A
general strategy for testing composite hypotheses is the so-called
{\em Generalized Likelihood Ratio Test} (GLRT). We suppose that the
set of alternatives is of the form $\lambda f$ where $\lambda$ is a
scalar and $f$ belongs to a subset $\cF$ of unit vectors of $\R^N$,
i.e. obeying $\|f\| = 1$ for all $f \in \cF$ (unless specified
otherwise, $\|\cdot\|$ is the usual Euclidean norm).  In other words,
the alternative consists of multiples of a possibly exponentially
large set of candidate signals. In this setup, the GLRT takes the form
\begin{equation}
  \label{eq:glrt}
  \max_{\lambda \in \R, f \in \cF} \frac{L(\lambda f; y)}{L(0; y)},
\end{equation}
where $L(\lambda f; y)$ is the likelihood of the data when the true
mean vector is $\lambda f$. In the case of additive white noise, a
simple calculation shows that the GLRT is proportional to 
\[
\max_{\lambda \in \R, f \in \cF} \,\,\, e^{-\|y - \lambda
    f\|^2/2\sigma^2} = \max_{f \in \cF} \,\,\, e^{-\|y - \<y,
    f\> f\|^2/2\sigma^2},
\]
since for a fixed $f \in \cF$, the likelihood is maximized for
$\lambda = \<y, f\>$. It then follows from Pythagoras' identity $\|y -
\<y, f\> f\|^2 = \|y\|^2 - |\<y, f\>|^2$ so that the GLRT is equivalent
to finding the solution to
\[
\max_{f \in \cF} \,\,\, |\<y, f\>|^2,
\]
and comparing this value with a threshold. 

\subsection{The Best Path statistic}
\label{sec:bestpath}

Supplied with a chirplet graph, a reasonable strategy would be to
consider the class of signals which can be rewritten as a
superposition of piecewise linear chirps
\[
f(t) = \sum_{v \in W} \lambda_v f_v(t),
\]
where $W$ is any path in the chirplet graph and $(\lambda_v)$ is any
family of scalars, and apply the GLRT principle. In this setup, the
GLRT is given by
\[
\max_W \,\,\, \max_{(\lambda_v)} \,\,\, e^{-\|y - \sum_{v \in W}
    \lambda_v f_v\|^2/2\sigma^2} = \max_W \,\,\, \max_{(\lambda_v)}
\,\,\, \prod_{v \in W} e^{-\|y_v - \lambda_v f_v\|^2/2\sigma^2}, 
\]
where for each $v = (I,\mu)$, $y_v$ is the vector $(y_t)_{t \in I}$,
i.e.~the portion of $y$ supported on the time interval $I$. Adapting
the calculations detailed above shows that the GLRT is then equivalent to
\begin{equation}
  \label{eq:naive-GLRT}
  \max_{W} \,\,\, \sum_{v \in W} |\<y, f_v\>|^2. 
\end{equation}
In words, the GLRT simply finds the path in the chirplet graph which
maximizes the sum of squares of the empirical coefficients.  The major
problem with this approach is that the GLRT will naively overfit the
data. By choosing paths with shorter chirplets, one can find chirplets
with increased correlations (one needs to match data on shorter
intervals) and as a result, the sum $\sum_{v \in W} |\<y, f_v\>|^2$
will increase. In the limit of tiny chirplets, $|\<y, f_v\>|^2 =
\|y_v\|^2$ which gives
\[
 \max_{W} \,\,\, \sum_{v \in W} |\<y, f_v\>|^2 = \|y\|^2,  
\]
and one has a perfect fit!  There is an analogy with model selection
in multiple regression where one improves the fit by increasing the
number of predictors in the model. Just as in model selection, one
needs to adjust the goodness of fit with the complexity of the fit.

Let $W$ be a fixed path of length $|W|$. Then under the null
hypothesis, $\sum_{v \in W} |\<y, f_v\>|^2$ is distributed as a
chi-squared random variable with $|W|$ degrees of freedom. Thus for
fixed paths, we see that the value of the sum of squares along the
path grows linearly with the length of the path. In some sense, the
same conclusion applies to the maximal path; i.e. the value of the sum
of squares along a path of a fixed size $\ell$ also grows
approximately linearly with $\ell$, with a constant of proportionality
{\em greater than 1}. An exact quantitative statement would be rather
delicate to obtain in part because of the inherent complexity of the
chirplet graph but also because it would need to depend on the special
chirplet discretization. We refer the reader to \cite{DetectPath}.

The above analysis suggests taking a test statistic of the form 
\begin{equation}
  \label{eq:global-test}
  Z^* = \max_W  \,\,\, \frac{\sum_{v \in W} |\<y, f_v\>|^2}{|W|}, 
\end{equation}
which may be seen as a perhaps unusual trade-off between the goodness
of fit and the complexity of the fit.  This is of course motivated by
our heuristic argument which suggests that under the null hypothesis,
the value of the best path of length $\ell$, that solution to
\begin{equation}
  \label{eq:Tl}
  T^*_\ell := \max_{|W| \le \ell} \,\,\, \sum_{v \in W} |\<y, f_v\>|^2, 
\end{equation}
grows linearly with $\ell$, and is well concentrated around its mean
by standard large deviation inequalities. In other words, with
$Z^*_\ell := T^*_\ell/\ell$, one would expect $Z^*_\ell$ to be about
constant under $H_0$, at least for $\ell$ sufficiently large. This
would imply that if we ignored paths of small length, one would
expect---owing to sharp concentration---$Z^* = \max_\ell Z^*_\ell$ to
be about constant under $H_0$. Therefore, a possible decision rule
might be to reject $H_0$ if $Z^*$ is large.

\begin{figure}
  \begin{center}
    \includegraphics[height=3.5in]{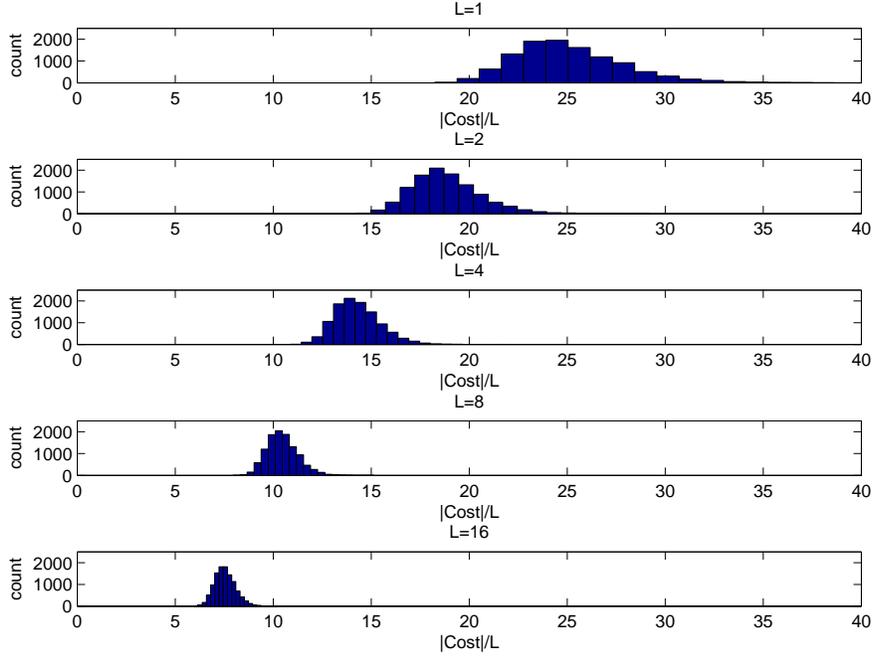}
  \end{center}
  \caption{\small Null distribution of $Z^*_\ell$ for values of $\ell$
    equal to 1, 2, 4, 8, 16. The mean and standard deviation are
    decreasing with $\ell$.}
  \label{fig:histograms}
\end{figure}

Numerical simulations confirm that under the null, $T^*_\ell$ grows
linearly with $\ell$ but they also show---as expected---deviations for
small values of $\ell$, see Figure \ref{fig:histograms}. For example,
with the discretization discussed in Section \ref{sec:discretization},
$\E Z^*_\ell$ seems to be decreasing with $\ell$. With this
discretization, $Z^*$ is also almost all the time attained with paths
of length 1 (one single chirplet) so that $Z^*$ is almost always equal
to $Z^*_1$. If we were to set a threshold based on the quantile of the
null distribution of $Z^*$ which basically coincides with that of
$Z_1^*$, we would lose some power to detect the alternative.  Suppose
indeed that there is signal. Then the signal may be strong enough so
that the observed value of $Z^*_\ell$ for some $\ell$ may very well
exceed the appropriate quantile of its null distribution, hence
providing evidence that there is signal, but too weak for the observed
$Z^*$ to exceed the appropriate quantile of its null
distribution. Hence, we would have a situation where we could in
principle detect the signal but would fail to do so because we would
use a low-power test statistic which is not looking in the right
place.

A more powerful approach in order to gather evidence against the null
consists in looking at the $Z^*_\ell$'s for many different values of
$\ell$, and find one which is unusually large.  Because we are now
looking at many test statistics simultaneously, we need a multiple
comparison procedure which would deal with issues arising in similar
situations, e.g. in multiple hypothesis testing \cite{MCP}.  For
example, suppose we are looking at $k$ values of $\ell$ and let
$q_\ell(\alpha)$ be the $\alpha$th quantile of the distribution of
$Z^*_\ell$. Then to design a test with significance level $\alpha$,
one could use the Bonferroni procedure and reject the null if one of
the $Z^*_\ell$'s exceeds $q_\ell(1-\alpha/k)$ (informally, one would
test each hypothesis at the $\alpha/k$ level).  The Bonferroni method
is known to be overly conservative in the sense that it has low power
and a better approach is to conduct an $\alpha$-level test is as follows:
\begin{enumerate}
\item Calculate the $p$-values for each of the $Z^*_\ell$ and find the
  minimum $p$-value.
\item Compare the observed minimum $p$-value with the distribution of
  the minimum $p$-value under the null hypothesis.
\end{enumerate}
In the first step, we are choosing the coordinate of the multivariate
test statistic that gives the greatest evidence against the null
hypothesis. In the second step, we compare our test statistic with
what one would expect under the null. We call this the {\em Best Path
  test/statistic} or the BP statistic for short. In Section
\ref{sec:simulations}, we will see that this simple way of combining
the information in all the coordinates of the multivariate test
statistic enjoys remarkable practical performance.

At this point, one might be worried that the computational cost for
calculating the $Z^*_\ell$'s is prohibitive. This is not the case.  In
fact, besides having sound statistical properties, the BP test is also
designed to be rapidly computable. This is the subject of Section
\ref{sec:algorithms}.


\subsection{Why multiscale chirplets?}
\label{sec:why}

If one were to use monoscale chirplets, i.e. a set of chirplets living
on time intervals of the form $[k 2^{-j}, (k+1) 2^{-j})$ for a {\em
  fixed} scale $2^{-j}$, then all the paths would have the same length
(equal to $2^j$) and the issue of how to best trade-off between the
goodness of fit and the complexity of the fit would of course
automatically disappear. One could then apply the GLRT
\eqref{eq:naive-GLRT}, which is rapidly computable via dynamic
programming as we will see in Section \ref{sec:sp}. 

Multiscale chirplets, however, provide a much richer structure.
Whereas a monoscale approach imposes to use templates of the same
length everywhere, the multiscale approach offers the flexibility to
use shorter templates whenever the instantaneous frequency exhibits a
complicated structure and longer templates whenever it exhibits a
simpler structure. In other words, the multiscale chirplet graph has
the advantage of automatically adapting to the unknown local
complexity of the signal we wish to detect. Moreover, with monoscale
models, one would need to decide which scale to use and this may be
problematic. The best scale for a given signal may not be the best for
a slightly different signal so that the whole business of deciding
upon a fixed scale may become rather arbitrary.  We are of course not
the first to advocate the power of multiscale thinking as most
researchers in the field have experienced it (the list of previous
`multiscale successes' is very long by now and ever increasing). Here,
we simply wish to emphasize that the benefits of going multiscale
largely outweigh the cost.

\section{Best Path Algorithms}
\label{sec:algorithms}

This section presents an algorithm for computing the Best Path
statistic, which requires solving a sequence of optimization problems
over all possible paths in the chirplet graph; for each $\ell$ in a
discrete set of lengths, we need to solve
\begin{equation}
  \label{eq:CSP}
  \max_W \,\,\, \sum_{v \in W} 
|\langle y, f_v \rangle|^2, \quad \text{ s.t. } \quad |W| \le \ell.  
\end{equation}
Although the number of paths in the graph is exponential in the sample
size $N$, the BP statistic is designed in such way that it allows the
use of network flow algorithms for rapid computation. We will find out
that the complexity of the search is of the order of the number of arcs in
the chirplet graph times the maximum length of the path we are willing
to consider. Later in this section, we will discuss proxies for the
Best Path statistic with even more favorable computational
complexities.

Before we begin, we assume that all the vertices in the chirplet graph
are labeled and observe that the chirplet graph is a directed and
acyclic graph, meaning that the vertices on any path in the graph are
visited only once, i.e. the graph contains no loops. Suppose that two
vertices $v$ and $w$ are connected, then we let $C(v,w)$ be the cost
associated with the arc $(v,w)$, which throughout this section is
equal to the square of the chirplet coefficient at the node $w$,
$C(v,w) = |\<y, f_w\>|^2$.  (We emphasize that nothing in the
arguments below depends on this assumption.) To properly define the
cost of starting-vertices, we could imagine that there is a
\emph{dummy vertex} from which all paths start and which is connected
to all the starting-vertices in the chirplet graph. We put $|E|$ and
$|V|$ to denote the number of arcs and vertices in the graph under
consideration.

\subsection{Preliminaries}
\label{sec:sp}

An important notion in graph optimization problems is that of {\em
  topological ordering}. A topological ordering of a directed acyclic
graph is an ordering of the vertices$(v_i)$, $i = 1, \ldots, |V|$,
such that for every arc $(v_i, v_j)$ of the graph, we have $i <
j$. That is, a topological ordering is a linear ordering of all its
vertices such that the graph contains an edge $(v_i, v_j)$, then $v_i$
appears before $v_j$ in the ordering. From now on, we will use the
notations $i$ or $v$ to denote vertices and $(i,j)$ or $(v,w)$ to
denote edges interchangeably.

Labeling chirplets in the chirplet graph is easy. We move along the
time axis from left to right, taking the smallest possible time step
(depending on the smallest allowable scale) and label all the
chirplets starting from the current position on the time axis; all
these chirplets are not connected to each other and, therefore, we may
order them freely. Any chirplet starting at a later time will receive
a larger topological label and, therefore, the chirplets are arranged
in topological order.

\newcommand{\pred}{\text{pred}}
Suppose we wish to find the so-called shortest path in the chirplet
graph, i.e.
\begin{equation}
  \label{eq:SP}
  \max_W \,\,\, \sum_{v \in W} |\<y, f_v\>|^2.  
\end{equation}
(In the literature on algorithms, this is called the shortest path
because by flipping the cost signs and interpreting the costs as
distances between nodes, this is equivalent to finding the path along
which the sum of the distances is minimum.) To find the shortest path,
one can use Dijkstra's algorithm which is known to be a good algorithm
\cite{Ahuja:NetworkFlows}. We let $i = 0$ be the source or dummy node
and $d(v)$ be the value of the maximum path from the source to node
$v$. Below, the array $\pred$ will be a list of the predecessor
vertices in the shortest path. That is, if $\pred(j)=i$, then the arc
$(i,j)$ is on the shortest path.

{\bf Algorithm for shortest path in a chirplet graph}. 
\begin{itemize}
\item Set $d(s)=0$ and $d(i) = 0$ for $i = 1, \ldots, |V|$. 
\item Examine the vertices in topological order. For $i = 1, \ldots, |V|$:
\begin{itemize}
\item Let $A(i)$ bet the set of arcs going out from vertex $i$.
\item Scan all the arcs in $A(i)$. For $(i,j)\in A(i)$, if $d(j) <
  d(i) + c(i,j)$, set $d(j) = d(i) + c(i,j)$ and $\pred(j)=i$.
\end{itemize}
\end{itemize}
Since every arc is visited only once, this shows that the maximum path
in the chirplet graph can be found in $O(|E|)$ where we recall that
$|E|$ is the number of edges in the graph. 

\subsection{The Best Path algorithm}

The idea of solving a Shortest Path Problem using an updated costs can
be used to solve a Lagrangian relaxation of the Constrained Shortest
Path Problem. This approach is well known in the field of Network
Flows. Solving the problem \eqref{eq:CSP} for every possible length
would give us the points defining the convex hull of the achievable
paths, i.e. the convex hull of the points $(|W|, C(W))$, where
$C(W)$ is the cost of the path $W$. A point on the convex hull is
solution to
\begin{equation}
  \label{eq:Lagrangian}
  \max_W \,\,\, \sum_{v \in W} |\<y, f_v \>|^2 - \lambda \, |W|, 
\end{equation}
where $\lambda$ is some positive number, which can be solved by the
Dijkstra's algorithm by setting $\tilde C(v,w) = C(v,w) -
\lambda$. Then one could try to solve a series of problems of this
type for different values of $\lambda$ to hunt down solutions of the
Constrained Shortest Path problem for different values of
length. There are many proposed rules in the literature for updating
$\lambda$ but nothing with guaranteed efficiency.

Perhaps surprisingly, although the Constrained Shortest Path Problem
is in general NP-complete for noninteger times, we can solve it in
polynomial time by changing the Shortest Path algorithm only slightly
\cite{Joksch:CSP}. 
We let $i = 0$ be the source node and $d(i,k)$ be the value of the
maximum path from the source to node $i$ using exactly $k$ arcs, where
$k$ ranges from 0 to $\ell_{\max}$. As before, we denote by
$\pred(i,\ell)$ the vertex which precedes vertex $i$ in the tentative
best path of length $\ell$ from the source node to vertex $k$. The
following algorithm solves the Constrained Shortest Path Problem in
about $O(\ell_{\max} \, |E|)$, where $\ell_{\max}$ is the maximum
number of vertices allowed in the path and $|E|$ is the number of
edges in the graph.
\paragraph{Best Path Algorithm}
\begin{itemize}
\item Set $d(0,\cdot) = 0$ and $d(i,\cdot)= 0$ for $i = 1, \ldots, |V|$. 
\item Examine vertices in topological order. For $i = 1, \ldots, |V|$:
\begin{itemize}
\item Let $A(i)$ bet the set of arcs going out from vertex $i$.
\item Scan arcs in $A(i)$. For $(i,j)\in A(i)$, for $k = 1, \ldots,
  \ell_{max}$, if $d(j,k) < d(i,k-1) + c(i,j)$, set $d(j,k) = d(i,k-1)
  + c(i,j)$ and $\pred(j,k)=i$.
\end{itemize}
\end{itemize}

This algorithm is slightly more expensive than the Shortest Path
algorithm since it needs to keep track of more distance labels. The
memory storage requirement is of size $O(|V| \times \ell_{\max})$ for
storing the distance labels and the predecessor vertices.  If we want
to include all possible lengths in \eqref{eq:CSP} so that $\ell_{\max}$
be of size about $N$ in the chirplet graph, then the memory would
scale as $O(N \times M_N)$ where $M_N$ is the number of chirplets.

\subsection{Variations}
\label{sec:variations}

There are variations on the BP statistic which have lower
computational costs and storage requirements, and this section
introduces one of them. Instead of computing \eqref{eq:CSP}, we 
could solve the Minimum-Cost-to-Time Ratio problem (MCTTR)
\begin{equation}
  \label{eq:MCTTR}
  \max_{W \in {\cal W}_k} \,\,\, \sum_{v \in W} 
\frac{|\<y, f_v\>|^2}{|W|}, 
\end{equation}
where for each $k$, ${\cal W}_k$ is a subset of all paths in the
chirplet graph. A possibility is to let $\mathcal{W}_0$ be the set of
all paths, $\mathcal{W}_1$ be the set of paths which cannot use
chirplets at the coarsest scale, $\mathcal{W}_2$ be the set of paths
which cannot use chirplets at the two coarsest scales, and so on.
Hence the optimal path solution to \eqref{eq:MCTTR} is forced to
traverse at least $2^k$ nodes. In this way, we get a family of near
optimal paths of various lengths.  There is an algorithm which allows
computing the MCTTR for a fixed $k$ by solving a sequence of Shortest
Path problems, see the Appendix. This approach has the benefit of
requiring less storage, namely, of the order of $O(|V|)$, and for each
$k$, the computational cost of computing the best path is typically of
size $O(|E|)$.

\section{Extensions}
\label{sec:extensions}

Thus far, we considered the detection problem of chirps with slowly
time-varying amplitude in Gaussian white noise and in this section, we
discuss how one can extend the methodology to deal with a broader
class of problems. 

\subsection{Colored noise}

We consider the same detection problem \eqref{eq:model} as before but
we now assume that the noise $z$ is a zero-mean Gaussian process with
covariance $\Sigma$. Arguing as in Section \ref{sec:teststatistic},
the GLRT for detecting an alternative of the form $\lambda f$ where
$\lambda \in \R$ and $f$ belongs to a class of normalized templates is
of the form
\[
\min_{\lambda \in \R, \, f\in \cF} \,\,\, e^{-(y - \lambda f)^T \Sigma^{-1} (y
- \lambda f)/2},
\]
which simplifies to 
\begin{equation}
  \label{eq:glrt2}
  \max_{f \in \cF} \,\,\, \frac{|y^T \Sigma^{-1} f|^2}{f^T \Sigma^{-1} f}. 
\end{equation}
Note that the null distribution of $|y^T \Sigma^{-1} f|^2/f^T
\Sigma^{-1} f$ follows a chi-square distribution with one degree of
freedom.

Our strategy then parallels that in the white noise model. We define
new chirplet costs by
\begin{equation}
  \label{eq:colored-cost}
  C(v) = \frac{|y^T \Sigma^{-1} f_v|^2}{f_v^T \Sigma^{-1} f_v}, 
\end{equation}
and compute a sequence of statistics by solving the Constrained
Shortest Path Problem
\begin{equation}
  \label{eq:newTl}
  T^*_\ell := \max_{W} \,\,\, \sum_{v \in W} C(v), \qquad |W| \le \ell.
\end{equation}
Note that we still allow ourselves to call such statistics $T^*_\ell$
since they are natural generalizations of those introduced earlier. We
then form the family $Z^*_\ell := T^*_\ell/\ell$ and find the Best
Path by applying the multiple comparison procedure of Section
\ref{sec:bestpath}. In short, everything is identical but for the cost
function which has been adapted to the new covariance structure. In
particular, once the new costs are available, the algorithm for
finding the best path is the same and, therefore, so is the
computational complexity of the search.

\subsection{Computation of the new chirplet costs}
\label{sec:compute-cost1}

In the applications we are most interested in, the noise process is
stationary and we will focus on this case. It is well known that the
Discrete Fourier Transform (DFT) diagonalizes the covariance matrix of
stationary processes so that
\[
\Sigma = F^* D F, \quad D = \text{diag}(\sigma_\omega^2), 
\]
where $F$ is the $N$ by $N$ DFT matrix, $F_{kt} = e^{-\i 2\pi k
  t/N}/\sqrt{N}$, $0 \le k, t \le N-1$, and $\sigma_1^2, \ldots,
\sigma_N^2$ are the eigenvalues of $\Sigma$.

To compute the chirplet costs, we need to evaluate the coefficients
$y^* \Sigma^{-1} f_v$. Observe that 
\[
y^* \Sigma^{-1} f_v = \tilde y^* f_v, \qquad \tilde y = \Sigma^{-1} y
= F^* D^{-1} F y.
\]
In other words, we simply need to compute $\tilde y$ and apply the
discrete chirplet transform. The cost of computing $\tilde y$ is
negligible since it only involves two 1D FFT of length $N$ and $N$
multiplications. Hence, calculating all the coefficients $y^*
\Sigma^{-1} f_v$ takes about the same number of operations as applying
the chirplet transform to an arbitrary vector of length $N$.

To compute the costs, we also need to evaluate $f_v^* \Sigma^{-1}
f_v$, which can of course be done offline. It is interesting to notice
that this can also be done rapidly. We explain how in the case where
the discretization is that introduced in Section
\ref{sec:discretization}.  First, observe that for any pair of
chirplets $f_v$, $f_{w}$ which are time-shifted from one another, we
have
\[
f_v^* \Sigma^{-1} f_v = f_w^* \Sigma^{-1} f_w 
\]
since $\Sigma^{-1}$ is time invariant. Thus we only need to consider
chirplets starting at $t = 0$. Second, letting $(\hat f[\omega])_{0
  \le \omega \le N-1}$ be the DFT of $(f[t])_{0 \le t \le N-1}$
\[
\hat f[\omega] = \frac{1}{\sqrt N} \sum_{t = 0}^{N-1} f[t] \, e^{-\i 2\pi \omega
  t/N}, 
\] 
we have that 
\[
f^* \Sigma^{-1} f = \sum_{\omega = 0}^{N-1} |\hat f[\omega]|^2/\sigma_\omega^2. 
\]
All the chirplets associated with the fixed time interval $[0,2^{-j})$
are of the form
\[
f_{a,b}[t] = |I|^{-1/2} \, e^{\i 2\pi(b t/N + a(t/N)^2/2)} \, 1_I(t), 
\] 
where $b = 0,1, \ldots, N -1$, and $a$ is a discrete set of slopes of
cardinality about $N/2^j$. Now the modulation property of the DFT
gives $\hat f_{a,b}[\omega] = \hat f_{a,0}[\omega - b]$ and so we only
need to compute the DFT of a chirplet with zero frequency offset. This
shows that for a fixed slope, we can get all the coefficients
corresponding to all offsets by means of the convolution
\[
f_{a,b}^* \Sigma^{-1} f_{a,b} = \sum_{\omega = 0}^{N-1} 
|\hat f_{a,0}[\omega-b]|^2/\sigma_\omega^2, 
\]
which can be obtained by means of 2 FFTs of length $N$. With the
assumed discretization, there are about $N/2^j$ slopes at scale
$2^{-j}$ and so computing $\hat f_{a,0}[\omega]$ for all slopes has a cost
of at most $O(N^2/2^j \cdot \log N)$ flops. Hence the total cost of
computing all the coefficients $f_v^* \Sigma^{-1} f_v$ is at most
$O(N^2 \log N)$ and is comparable to the cost of the chirplet
transform.

\subsection{Varying amplitudes}

We are still interested in detecting signals of the form $S(t) = A(t)
\exp(\i \lambda \varphi(t))$ but $A(t)$ is such that fitting the data
with constant amplitude chirplets may not provide local correlations
as large as one would wish; i.e.  one would also need to adjust the
amplitude of the chirplet during the interval of operation.

To adapt to this situation, we choose to correlate the data with
templates of the form $p(t) \, e^{\i \varphi_v(t)} 1_{I}(t)$, where
$p(t)$ is a smooth parametric function, e.g. a polynomial of a degree
at most 2, and $e^{\i \varphi_v(t)} 1_{I}(t)$ is an unnormalized
  chirplet. The idea is of course to look for large correlations with
  superpositions of the form
\[
\sum_{v \in W} p_v(t) \tilde f_v(t), \quad \tilde f_v(t) = e^{\i
  \varphi_v(t)} \, 1_I(t).
\]
Fix a path $W$. In the white noise setup, we we would select the
individual amplitudes $p_v$ to minimize
\begin{equation}
\label{eq:trivial}
  \sum_{v \in W} \sum_{t \in I} |y_v(t) - p_v(t) \tilde f_v(t)|^2,   
\end{equation}
and for each chirplet, $p_v$ would be adjusted to minimize $\sum_{t
  \in I} |y_v(t) - p_v(t) \tilde f_v(t)|^2$.  Put $\tilde y_v(t) =
y_v(t) e^{\i \varphi_v(t)}$ and let $P$ denote the projector onto a
small dimensional subspace $S$ of smooth functions over the interval
$I$, e.g. the space of polynomials of degree 2; if $b_1(t), \ldots,
b_k(t)$ is an orthobasis of $S$, then $P^*$ is the matrix with the
$b_i$'s as columns. The minimizer $p_v$ is then given by $P \tilde
y_v$ and it follows from Pythagoras' identity that $\|\tilde y_v -
P\tilde y_v\|^2 = \|\tilde y_v\|^2 - \|P \tilde y_v\|^2$. We introduce
some matrix notations and let $\Phi_v = \text{diag}(\e^{\i
  \varphi_v(t)})$ so that $\tilde y_v = \Phi_v^* y_v$.  Then one can
apply the same strategy as before but with chirplet costs equal to
\begin{equation}
  \label{eq:cost-amplitude1}
  C(v) = \|P
  \tilde y_v\|^2 = \|A_v \, y\|^2, \qquad A_v = P \Phi_v^*.
\end{equation}
It follows from this equation that the complexity of computing these
costs is of the same order as that of computing the chirplet
transform.

Suppose now that the covariance is arbitrary, then one chooses $p_v$
solution to
\[
\min_{p \in S} \,\,\, (y - \Phi_v p)^* \Sigma^{-1} (y - \Phi_v p) =
y^* y - y^* \Sigma^{-1} A^* (A \Sigma^{-1} A^*)^{-1} A \Sigma^{-1} y,
\]
so that the general chirplet cost is of the form 
\begin{equation}
  \label{eq:cost-amplitude2}
  C(v) =  y^* \Sigma^{-1} A^* (A \Sigma^{-1} A^*)^{-1} A \Sigma^{-1} y, 
\qquad A_v = P \Phi_v^*.
\end{equation}

\subsection{Computing the general chirplet costs}
\label{sec:compute-cost2}

We briefly argue that the number of flops needed to compute all the
costs \eqref{eq:cost-amplitude2} is of the same order as that needed
for the original chirplet transform. Rewrite the cost
\eqref{eq:cost-amplitude2} as
\[
C(v) = x_v^* B_v^{-1} x_v \qquad x_v = A_v \Sigma^{-1} y, \quad B_v = A_v
\Sigma^{-1} A_v^*.
\]
Then all the $x_v$'s and all the $B_v^{-1}$'s can be calculated
rapidly. Once $x_v$ and $B_v$ are available, computing $x_v B^{-1}
x_v$ is simply a matter of calculating $B_v^{-1} x$---either a small
matrix multiplication or the solution to a small linear system
depending on whether we store $B_v$ or $B_v^{-1}$---followed by an
inner product.

We begin with the $x_v$'s.  We have already shown how to apply
$\Sigma^{-1}$ rapidly by means of the FFT, see Section
\ref{sec:compute-cost1}. With $\tilde y = \Sigma^{-1} y$, the $j$th
coordinate of $x_v$ is given by
\[
\sum_t \tilde y(t) b_j(t) \overline{f_v(t)}.  
\]
We then collect all the $x_v$'s by multiplying the data with the
appropriate basis functions and taking a chirplet transform.  If
we have $k$ such basis functions per interval, the number of flops
needed to compute all the $x_v$'s is about $k$ times that of the
chirplet transform.

We now study $B_v$. Note that for each $v$, $B_v$ is a Hermitian $k$
by $k$ matrix and so that we only need to store $k(k+1)/2$ entries per
chirplet; e.g. 3 in the case where $k = 2$, or 6 in the case where $k
= 3$. Also in the special case where $k = 1$ (constant amplitude), $P$
is the orthogonal projection onto the constant function equal to one
and $B_v = n_I^{-1} (f_v^* \Sigma^{-1} f_v)$, where $n_I$ is the number
of discrete points in the interval $I$. Computing $B_v$ is nearly
identical to computing $f_v^* \Sigma^{-1} f_v$, which we already
addressed. First, by shift invariance, we only need to consider
chirplet indices starting at time $t = 0$. Second, we use the diagonal
representation of $\Sigma^{-1}$ to write the $(i,j)$ entry of $B_v$
as
\[
\sum_{\omega = 0}^{N-1} \widehat{f_v b_i}[\omega] \, 
\overline{\widehat{f_v b_j}[\omega]} \, \sigma_\omega^{-2}. 
\]
Two chirplets $f_v$ and $f_w$ at the same scale and sharing the same
chirprate differ by a frequency shift $\omega_0$ so that $\widehat{f_w
  b_\ell}[\omega] = \widehat{f_v b_\ell}[k - \omega_0]$. Again, one
can use circular convolutions to decrease the number of
operations. That is, we really only need to evaluate $B_v$ for
chirplets starting at $t = 0$ and with vanishing initial frequency
offset. In conclusion, just as in the special case and for the
discretization described in Section \ref{sec:discretization}, one can
compute all the $B_v$'s in order $O(N^2 \log N)$ flops. To be more
precise, the cost is here about $k(k+1)/2$ that of computing $f_v^*
\Sigma^{-1} f_v$ for all chirplets.

\section{Numerical Simulations}
\label{sec:simulations}

We now explore the empirical performance of the detection methods
proposed in this paper. To this end, we have developed {\em ChirpLab},
a collection of Matlab routines that we have made publicly available,
see Section \ref{sec:chirplab}.  For simplicity, we use a chirplet
dictionary with the discretization discussed in Section
\ref{sec:discretization}. We also consider a slightly different
chirplet graph which assumes less regularity about the instantaneous
frequency of the unknown chirp; namely, two chirplets are connected if
and only if they live on adjacent time intervals and if the
instantaneous frequencies at their juncture coincide. In practical
situations such as gravitational wave detection, the user would be
typically given prior information about the signal she wishes to
detect which would allow her to fine-tune both the discretization and
the connectivities for enhanced sensitivity. We will discuss these
important details in a separate publication. Our goal here is merely
to demonstrate that the methodology is surprisingly effective for
detecting a few unknown test signals.

\subsection{The basic setup}

\newcommand{\vf}{\varphi}

\begin{figure}
\begin{center}
 \begin{tabular}{ccc}
      \includegraphics[scale=0.4]{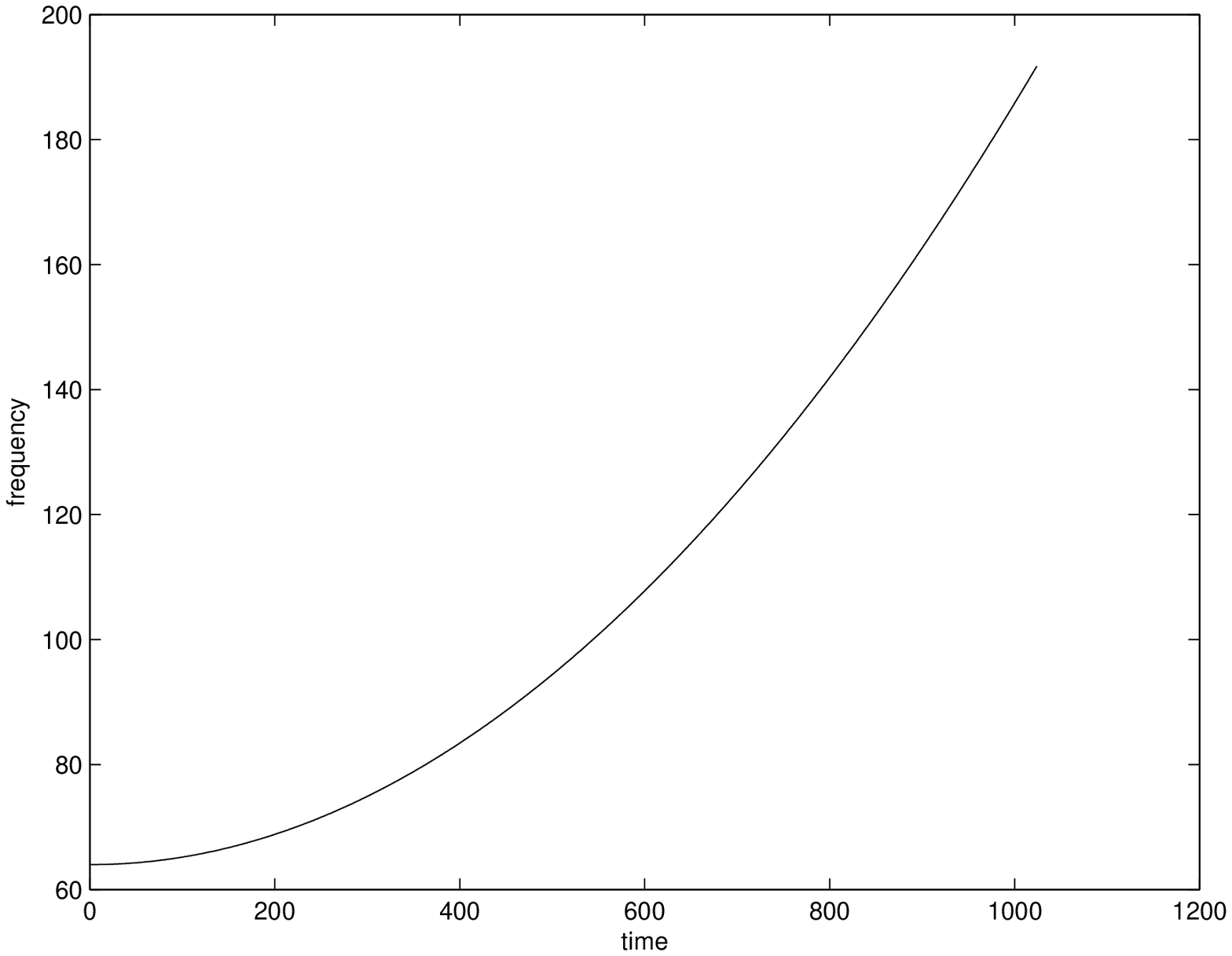} & & 
      \includegraphics[scale=0.4]{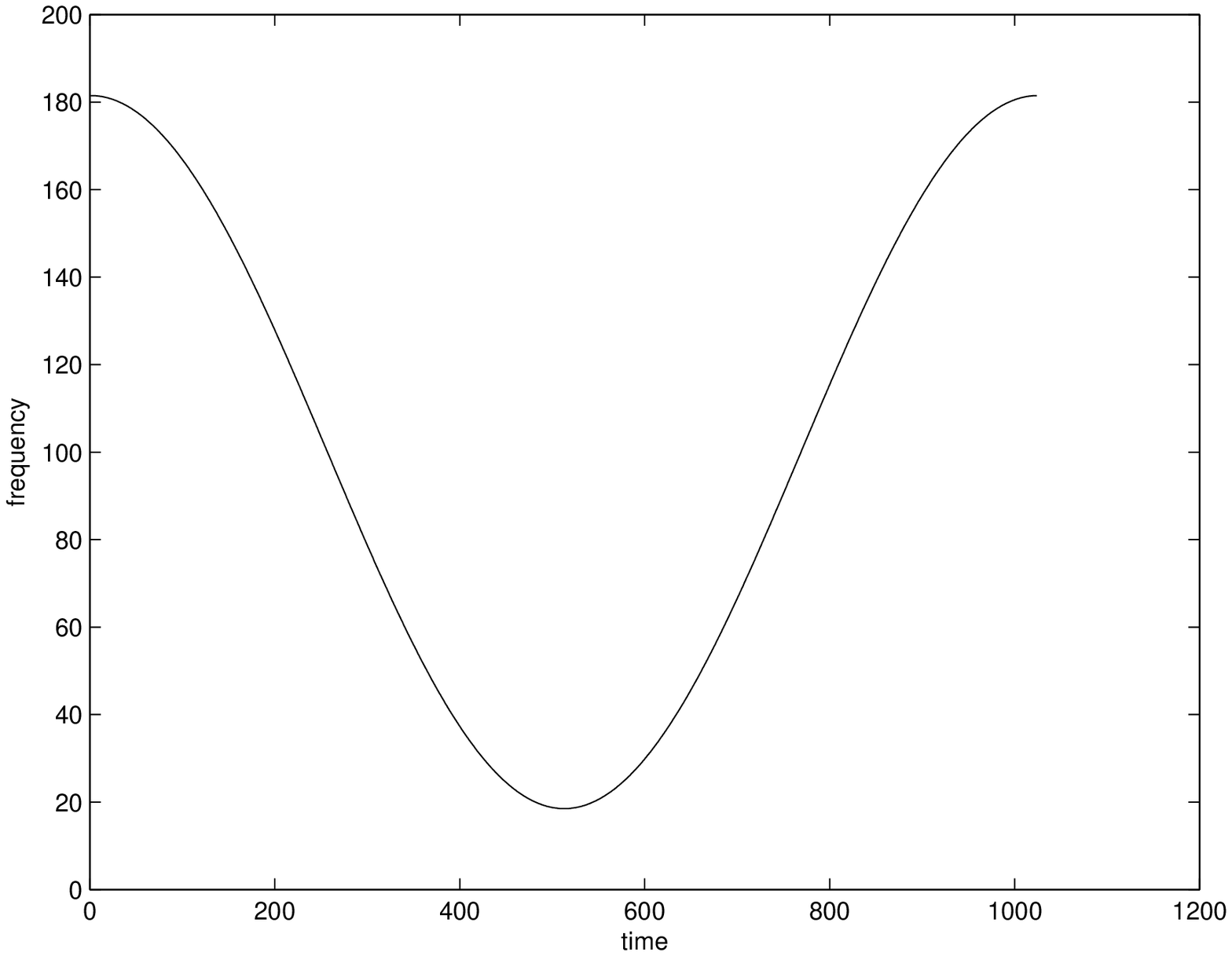}\\ 
(a) & & (b)
    \end{tabular}
\end{center}
\caption{\small Instantaneous frequency $\vf'(t)$ of the chirps under study. 
(a) Cubic phase chirp. (b) Cosine phase chirp.}
\label{fig:instfreq}
\end{figure}

\begin{figure}
\begin{center}
 \begin{tabular}{ccc}
\includegraphics[scale=0.4]{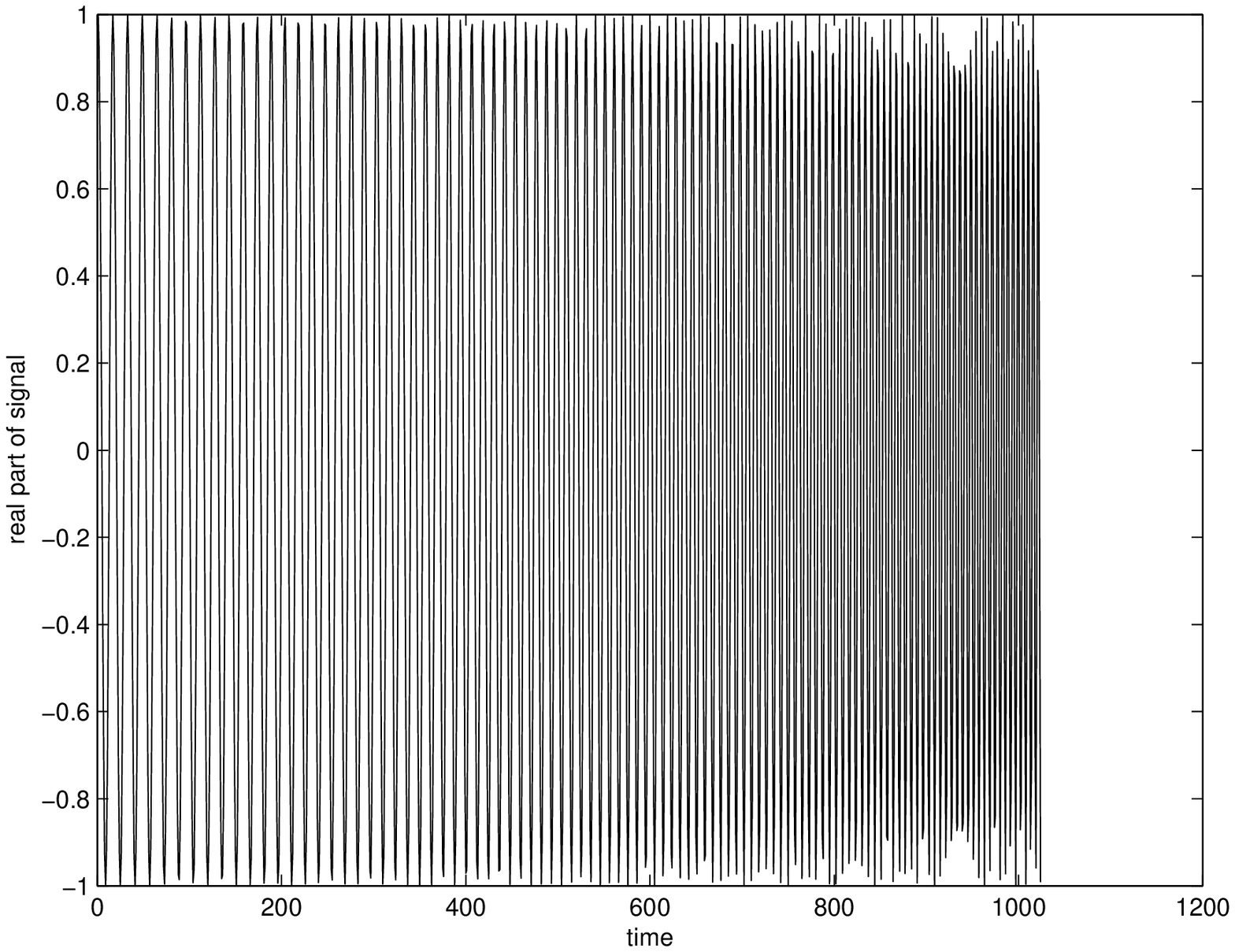} & & 
\includegraphics[scale=0.4]{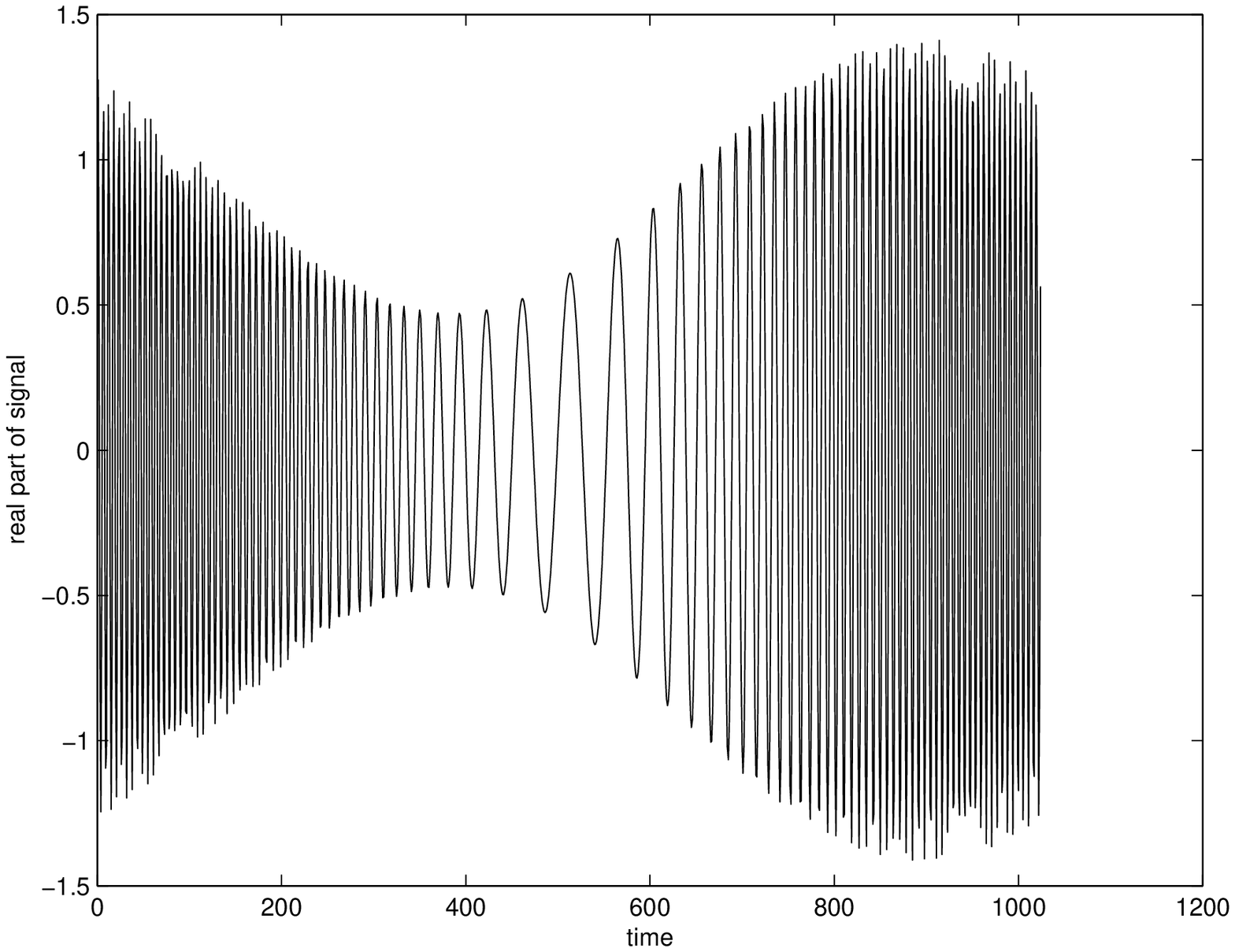}\\
(a) & & (b)
    \end{tabular}
\end{center}
\caption{\small Real part $A(i/N) \cos (N\vf(i/N))$, $i = 0, \ldots,
  N-1$ of the of the chirps under study.  (a) Cubic phase chirp. (b)
  Cosine phase chirp. The cosine phase chirp has a slowly varying
  amplitude. Note that the instantaneous frequency depends on the
  sample size $N$.}
\label{fig:signals-realpart}
\end{figure}

We generated data of the form 
\begin{equation*}
y_i = \alpha S_i + z_i, \qquad i = 0,1,\ldots,N-1;
\end{equation*}
where $(S_i)$ is a vector of equispaced time samples of a
complex-valued chirp, and where $(z_i)$ is a complex-valued white
noise sequence: $z = z^0 + \i \, z^1$ where $z^0$ and $z^1$ are two
independent vectors of i.i.d. $N(0,1/2)$ variables. Note that $E
|z_i|^2 = 1$ and $E \|z\|^2 = N$. In this setup, we define the SNR as
the ratio
\begin{equation}
  \label{eq:SNR}
  \text{SNR} = \frac{\|\alpha S\|}{\sqrt{N}}. 
\end{equation}
We have chosen to work with complex-valued data and want to emphasize
that we could just as well perform simulations on real-valued data and
detect real-valued signals, see the Appendix for details.  In all our
experiments, the signal $S$ obeys the normalization $\|S\| = \sqrt{N}$
so that the parameter $\alpha$ actually measures the SNR.  We
considered signals of size $N = 512, 1024, 2048, 4096$. The chirps are
of the form
\begin{equation}
  \label{eq:basic-chirp}
  S(t) = A(t) e^{i N \varphi(t)}, 
\end{equation}
and sampled at the equispaced points $t_i = i/N$, $i = 0, 1, \ldots,
N-1$. We considered two test signals.
\begin{enumerate}
\item A {\em cubic phase chirp} with constant amplitude:
\[
A(t) = 1, \qquad \varphi(t) = t^3/24 + t/16. 
\] 
\item A {\em cosine phase chirp} with slowly varying amplitude:
\[
A(t) = 2 + \cos(2\pi t + \pi/4), \qquad \varphi(t) = 
\sin(2 \pi t)/4\pi + 200\pi t/1024.
\] 
\end{enumerate}
Note that because of the factor $N$ in the exponential
\eqref{eq:basic-chirp}, we are not sampling the same signal at
increasingly fine rates. Instead, the instantaneous frequency of $S$
is actually changing with $N$ and is equal to $N \varphi'(t)$ so that
the signal may oscillate at nearly the sampling rate no matter what
$N$ is.  Figures~(\ref{fig:instfreq}) and (\ref{fig:signals-realpart})
show the rescaled instantaneous frequency, $\varphi^{\prime}(t)$ and the
real part of the signals under study for $N=1024$.

For detection, we use the BP test statistic introduced in Section
\ref{sec:bestpath} with $\{1,2,4,8,16\}$ as our discrete set of path
lengths.  We estimated the distribution of the minimum $P$-value under
the null hypothesis via Monte Carlo simulations, and selected a
detection threshold giving a probability of false detection (Type I
error) equal to $5\%$ (.05 significance level). 
\begin{itemize}
\item For each signal length, we randomly sampled about 10,000
  realizations of white noise to compute the 5\% detection level (the
  quantile of the minimum $P$-value distribution).

\item For each signal length, each signal and each SNR, we sampled the
  data model about 1,000 times in order to compute detection rates, or
  equivalently the so-called power curves.
\end{itemize}

In these simulations, we only considered chirplets with positive
frequencies and for the larger signal sizes, $N=2048, 4096$, we
restricted ourselves to discrete frequencies on the interval
$\{0,\ldots,N/4-1\}$ to save computational time.  In all cases the
slope parameters $a_\mu$ of the chirplets (see equation
\eqref{eq:chirplet}) ranged from $-\pi N$ to $\pi N$ with a
discretization at scale $2^{-j}$ of the form $a_\mu = 2\pi N (-1/2 + k
\cdot m 2^{j-J})$ where $J = \log_2 N$; $m=1$, $k\in \{0,\dots,
2^{J-j} \}$ for signal lengths $N = 512, 1024, 2048$ and $m = 4$, $k
\in \{0,\dots, 2^{J-j-2} \}$ for $ N= 4096$. This ensures that any
endpoint of a dyadic interval is an integer multiple of $2\pi$. 

The scales considered ranged from the coarsest $2^0$ to $2^{-s}$ with
$s=6$ for $N=512,1024$, $s=5$ for $N=2048$ and $s=4$ for $N=4096$ (the
motivation is again speed). In practice, these parameters would depend
upon the application and would need to be selected with care.  Tables
\ref{table:cosinecorrelation} and \ref{table:cubiccorrelation} show
the correlation between the waveforms and the best chirplet path with
a fixed length.  Although we use a coarser discretization and fewer
scales when $N = 4096$, the correlation is still very high, at least
for path lengths 8 and 16. Table \ref{table:cosinecorrelationwithamp}
shows the correlations between the cosine phase chirp and chirplets
with adapted amplitudes.  As expected, the correlation increases.
\begin{table}
\begin{center}
\begin{tabular}{|c|c|c|c|c|c|}
\hline
 signal length $N$ & $\ell=1$ & $\ell=2$ & $\ell=4$  & $\ell=8$ & $\ell=16$ \\
\hline
512 & 0.0718 & 0.4318 & 0.7126 &  0.9905 &  0.9982 \\
1024 & 0.0453 & 0.2408 & 0.5784 & 0.9814 & 0.9981 \\
2048 & 0.0306 & 0.1643 & 0.5107 & 0.9469 & 0.9976 \\
4096 & 0.0229 &    0.0953  &  0.4265  &  0.8158  &  0.9917 \\
\hline
\end{tabular}
\end{center}
\caption{\small Correlations between the cosine phase signal and the
  best chirplet path with fixed lengths $\ell \in \{1,2,4,8,16 \}$.}
\label{table:cosinecorrelation}
\end{table}

\begin{table}
\begin{center}
\begin{tabular}{|c|c|c|c|c|c|}
\hline
signal length N & $\ell=1$ & $\ell=2$ & $\ell=4$  & $\ell=8$ & $\ell=16$ \\
\hline
512 & 0.2382  & 0.8733 & 0.9903 & 0.9979 & 0.9999 \\
1024 & 0.1498 & 0.6575 & 0.9883 & 0.9985 & 0.9997 \\
2048 & 0.0932 & 0.3836 & 0.9671 & 0.9976 & 0.9995 \\
4096 & 0.0590  &  0.2373 &  0.8734 &  0.9903 &  0.9971 \\
\hline
\end{tabular}
\end{center}
\caption{\small Correlations between the cubic phase signal and the best
  chirplet path with fixed lengths $\ell \in \{1,2,4,8,16 \}$.}
\label{table:cubiccorrelation}
\end{table}

\begin{table}
\begin{center}
\begin{tabular}{|c|c|c|c|c|c|}
  \hline
  $d$ : degree of polynomials & 
$\ell=1$ & $\ell=2$ & $\ell=4$  & $\ell=8$ & $\ell=16$ \\
  \hline
  $[2, 1, 1, 1, 1, 1]$ & 0.1481 & 0.2852 & 0.5699 & 0.9612 & 0.9995 \\
  $[2, 2, 2, 1, 1, 1]$ & 0.1481 & 0.3337 & 0.5999 & 0.9612 & 0.9995 \\
  $[2, 2, 2, 2, 2, 2]$ & 0.1481 & 0.3337 & 0.6122 & 0.9823 & 0.9999 \\
  \hline
\end{tabular}
\end{center}
\caption{\small Correlations between the cosine phase signal and the
  best chirplet path with fixed lengths $\ell \in \{1,2,4,8,16\}$
  (chirplets with varying amplitude). $N=2048$.  The first column
  indicates the degree of the polynomial used to fit the
  amplitude. The entry $d_j$ in $d=[d_0,d_1,\ldots,d_5]$ is the degree
  of the polynomial at scale $2^{-j}$.}
\label{table:cosinecorrelationwithamp}
\end{table}

\subsection{Results from simulations}

To measure the performance of the BP statistic, we fix the probability
of Type I error at $5\%$ and estimate the detection rate, the
probability of detecting a signal when there is signal. We compute
such detection curves for various SNRs \eqref{eq:SNR}.  To limit the
number of computations we focus on a small set of signal levels around
the transition between a poor and a nearly perfect detection.

\begin{figure}[h]
\centerline{
\includegraphics[scale=0.5]{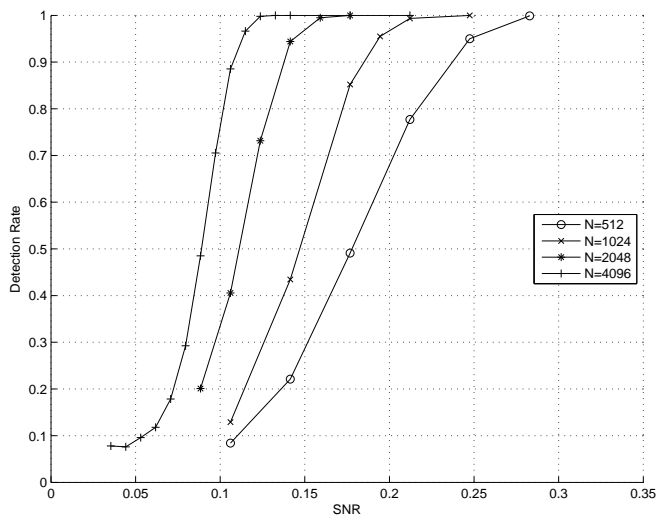}
}
\caption{\small Detection rates of the cubic phase chirp with the BP
  method. The type I error is fixed at 5\%.}
\label{fig:detrate-cubicphase}
\end{figure}

\begin{figure}[h]
\centerline{
\includegraphics[scale=0.5]{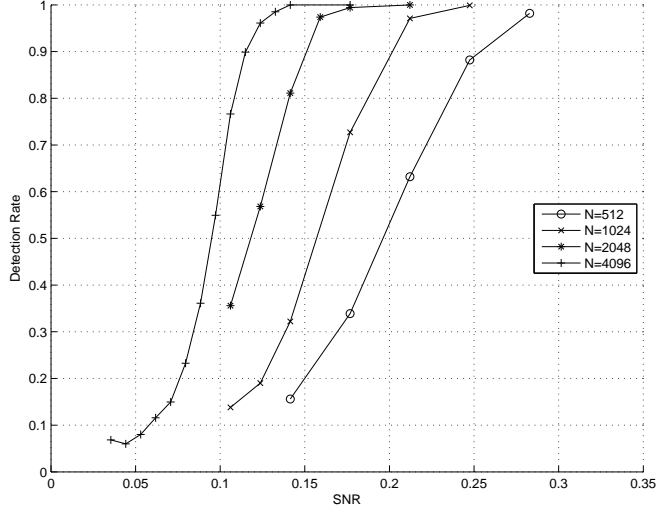}
}
\caption{\small Detection rates of the cosine phase chirp with the BP
  method. The type I error is fixed at 5\%.  }
\label{fig:detrate-cosphase}
\end{figure}

Figures \ref{fig:detrate-cubicphase} and \ref{fig:detrate-cosphase}
present results of a simulation study and display the power curves for
both chirps and for various sample sizes. Of course, as the sample
size increases, so does the sensitivity of the detector (even though
the signal is changing with the sample size). We also note that the
detection of the cubic phase chirp is slightly better than that of the
cosine phase chirp which was to be expected since the cubic phase
chirp is slightly less complex. (Simulations where one also adapts the
amplitude give similar results.)  

Consider the cosine phase chirp with time-varying amplitude and a
sample size $N$ equal to $4,096$. Then the SNR for a detection level
in the $95\%$ range is about .12. This means that one can reliably
detect an unknown chirp of about this complexity when the amplitude of
the noise is about 8 times that of the unknown signal.

\begin{figure}[h]
\centerline{
\includegraphics[scale=0.5]{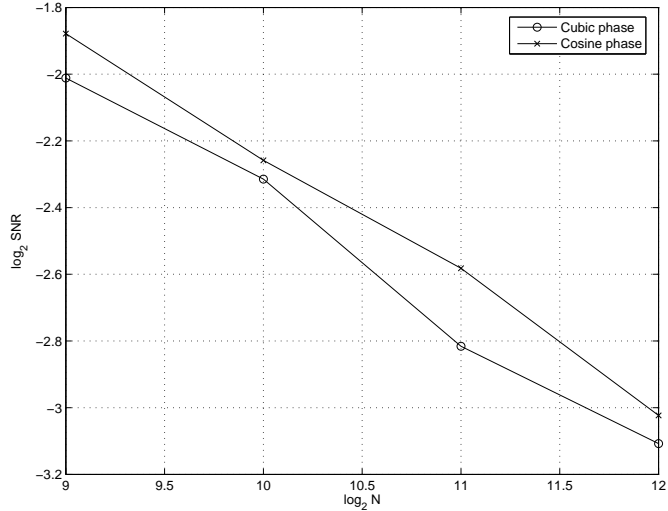}
}
\caption{\small Log-log (base-2) plot of the estimated SNR (for both
  chirps) at the $95\%$ detection rate versus signal length $N$.
  Again the Type I error is fixed at $5\%$. In both cases, the slope
  is approximately equal to -0.4.}
\label{fig:SNRvsN-log2plot}
\end{figure}

It is interesting to study the performance gain when we increase the
signal length. Fix a detection rate at 95\% and plot the SNR that
achieves this rate against the sample size $N$. Figure
\ref{fig:SNRvsN-log2plot} shows the base-2 logarithm of the estimated
SNR (using a simple linear interpolation of the power curves) versus
the logarithm of the sample size. The points roughly lie on a line
with slope -0.4; as we double the signal length from $N$ to $2N$, the
SNR required to achieve a 95\% detection rate is about $2^{-0.4}
\approx 0.76$ times that required to achieve the same detection rate
for the signal length $N$. In a parametric setting, we would
asymptotically expect a slope of -0.5. The fact that the slope is
slightly higher than this is typical of nonparametric detection
problems which deal with far richer classes of unknown signals
\cite{Ingster:NonparamDetectionBook}.

\subsubsection{Comparison with the detection a known signal}

In order to see how sensitive our test statistic really is, it might
be instructive to compare the detection rates with those one would
achieve if one had full knowledge about the unknown signal. We then
consider a simple alternative
\[
H_1 : y = \alpha S_0 + z,  
\] 
where the signal is known. 
That is, if there is signal, we know {\em exactly} how it looks
like. The standard likelihood ratio test (LRT) gives the optimal test
in terms of maximizing the power of detection at a given confidence
level. A simple calculation shows that the 5\% level, the power
function of the LRT is equal to $\Phi(1.65 - \text{SNR} \sqrt{2N})$
where $\Phi$ is the cumulative distribution of a standard
normal. Figure \ref{fig:linearChirpDetRates} shows this power curve
together with those obtained via the BP test for a sample size $N =
4096$. The horizontal gap between curves indicates the ratio between
SNRs to achieve the same detection rate. Consider a detection level
equal to about $95\%$.  Our plot shows that one can detect a
completely unknown signal via the BP statistic with the same power
than that one would get by knowing the signal {\em beforehand}
provided that the amplitude is about 3 times as large. Note that this
ratio is small and may be thought as the price one has to pay for not
knowing in advance what it is.

\subsubsection{Detection of a monochromatic sinusoid}

To appreciate the performance of the BP statistic, it might be a good
idea to study a more subtle problem. Suppose that the unknown signal
is a monofrequency sinusoid. If there is signal, we know it is of the
form $S(t) = e^{i\omega t + \phi}$, where the frequency $\omega$ and
the phase shifts are unknown. Consider the simpler case where for a
discrete signal of length $N$, $\omega \in \{0, \ldots, N - 1\}$ is
one of the $N$ Nyquist frequencies. Letting $y^0$ and $y^1$ be the
real and imaginary parts of the data $y$, the GLRT would maximize
\[
\sum_{0 \le t \le
  N-1} y^0_t \cos(2\pi kt /N + \phi) + y^1_t
\sin(2\pi kt /N + \phi),
\]
over $k = 0, 1, \ldots, N-1$ and $\phi \in [0, 2\pi]$. One can take
the maximum over $\phi$ and check that the GLRT is equivalent to
maximizing
\[
\left| \sum_{0 \le t \le N-1} y_t  e^{-i2\pi k t/N} \right| 
\]  
over $k$.  Thus, the GLRT has a simple structure. It simply computes
the DFT of the data, and compares the maximal entry of the response
with a threshold. (Note the resemblance of this problem with the
famous problem of testing whether the mean of Gaussian vector is zero
versus an alternative which says that one of its component is
nonzero.)


We could also make the problem a tiny bit harder by selecting the
frequency arbitrarily, i.e. not necessarily a multiple of $2\pi$ but
anything in the range $[0,2\pi N]$. In this case, the method described
above would be a little less efficient since the energy of the signal
would not be concentrated in a single frequency mode but spill over
neighboring frequencies. The GLRT would ask to correlate the data with
the larger collection of monofrequency signals which in practice we
could approximately achieve by oversampling the DFT (e.g. we could
select a finer frequency discretization so that the correlation
between the unknown monochromatic signal we wish to detect and the
closest test signal exceeds a fixed tolerance, e.g. .90 or .99).  

\begin{figure}[h]
\centerline{
\includegraphics[scale=0.5]{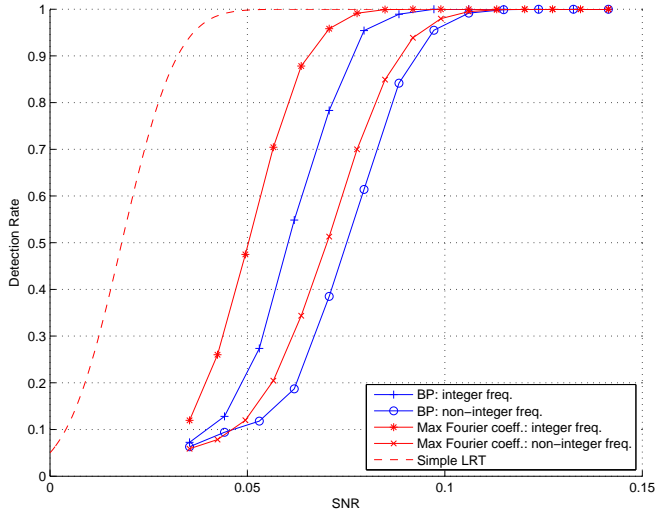}
}
\caption{\small Comparison of the BP and GLRT (based on the maximum
  modulus of the Fourier coefficients) for monochromatic
  sinusoids. The type I error is set at $5\%$}
\label{fig:monoFreqDetRates}
\end{figure}

We compare the detection rate curve for detecting (i)  a
monochromatic sinusoid with integer frequency and (ii) a monochromatic
sinusoid with arbitrary frequency using the maximum absolute DFT
coefficient on one hand, and the BP test on the other hand.  The
signals in (i) and (ii) are equispaced samples from $S_1(t) = e^{\i
  2\pi \frac{N}{8}t}$ and $S_2(t) = e^{\i 2\pi(\frac{N}{8} +
  \frac{1}{2})t}$. The signal length $N$ is equal to 4096.  Figure
\ref{fig:monoFreqDetRates} displays the detection rates.  Consider the
95\% detection rate. Then for (i) the SNR for the BP test is about
$20\%$ higher than that for the GLRT. In (ii) the SNR is only $8\%$
higher.  Also, at this detection level, the ratio between the SNRs for
the cosine phase chirp and the monofrequency is about 1.75.  These
results show that `the price we pay' for being adaptive and having the
ability to detect a rich class of chirping signals is low.

\subsubsection{Detection of a linear chirp}

\begin{figure}[h]
\centerline{
\includegraphics[scale=0.5]{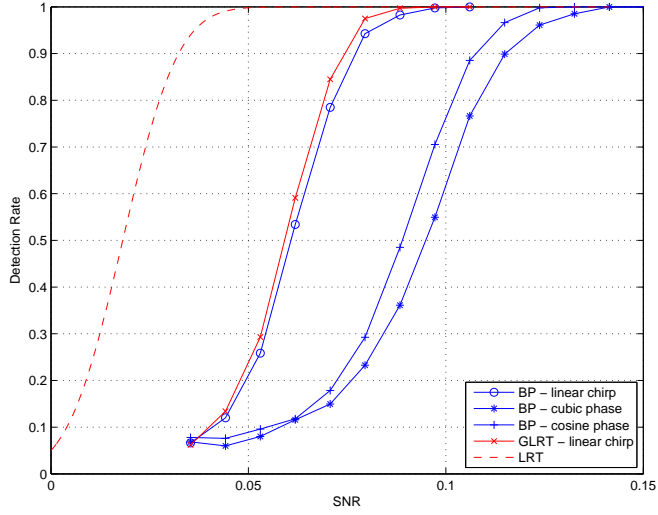}
}
\caption{\small Comparison of the BP and GLRT detection rates over a
  set of linear chirps. The type I error is set at $5\%$. Detection
  rates are plotted along with the detection rates for the cubic and
  cosine phase chirps.}
\label{fig:linearChirpDetRates}
\end{figure}

To study `the price of adaptivity,' we also consider the problem of
detecting linear chirps. Suppose that the unknown signal are sampled
values of a linear chirp of the form $S(t) = e^{i 2 \pi N \varphi(t)}$
where $\varphi(t) = a t^2/2 + b t + c$.  Here, $N = 4096$ and the
coefficients $a,b,c$ are adjusted so that the unknown linear chirp is
a complex multiple of a chirplet at the coarsest scale (the GLRT is
then the BP test restricted to paths of length 1).
In the simulations, we selected a chirp with $a = 1/8$, $b = 1/16$,
and $c = 0$ so that the instantaneous frequency $N\varphi'(t)$ increased
linearly from 256 to 768. Figure \ref{fig:linearChirpDetRates}
displays the detection rates for the GLRT and the BP test with $\{1,
2, 4, 8, 16 \}$ as path lengths.  The detection
rates for the BP test and the GLRT are almost the same; the ratio
between the SNRs required to achieve a detection rate of about $95\%$
is about 1.05. This shows the good adaptivity properties of the BP
test.  For information, the plot also shows that one can detect a
completely unknown signal via the BP statistic with the same power
than that one would get for detecting {\em a linear chirp} via the
GLRT provided that the amplitude is about 1.5 times as large.

\subsection{Empirical adaptivity on a simulated gravitational wave}

\begin{figure}[h]
\centerline{
\includegraphics[scale=0.5]{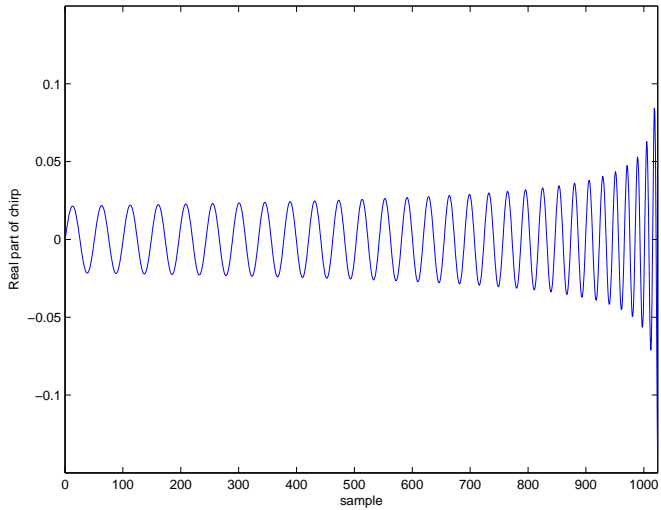}
}
\caption{\small Real part of a simulated gravitational wave}
\label{fig:gwsignal_realpart}
\end{figure}

Earlier, we argued that the GLRT or the method of matched filters
would need to generate exponentially many waveforms to provide good
correlations with the unknown signal of interest. The idea underlying
the chirplet graph is that one can get very good correlations by
considering a reasonably sized dictionary and considering correlations
with templates along a path in the graph. Figure
\ref{fig:gwsignal_realpart} shows the real part of a `mock'
gravitational waveform whose instantaneous frequency and amplitude
increase roughly as a power law as in Section \ref{sec:gw}.  The
waveform is $S(t) = A(t) e^{\i \varphi(t)}$ where the phase is 
\[
\varphi(t) = a_0 (t_c - t)^{5/8} + a_1 (t_c - t)^{3/8}
+a_2 (t_c - t)^{1/4} +  a_3(t_c - t)^{1/8}, 
\]
and the amplitude is given by $A(t) = [\varphi' (t)]^{2/3}$
(see Figure \ref{fig:gwsignal_realpart}). 
The coefficients $a_0,\ldots,a_3$ where chosen from the post-Newtonian
approximation for a binary inspiral as described in
\cite{anderson:tfgwdetection,allen}. The coefficient $t_c$ is the time
of coalescence. The masses of the two bodies where both chosen to be
equal to $14$ solar masses and the sampling rate was $2048$ Hz. We
studied the last 1024 samples of the waveform.

\begin{figure}[h]
\centerline{
\includegraphics[scale=0.8]{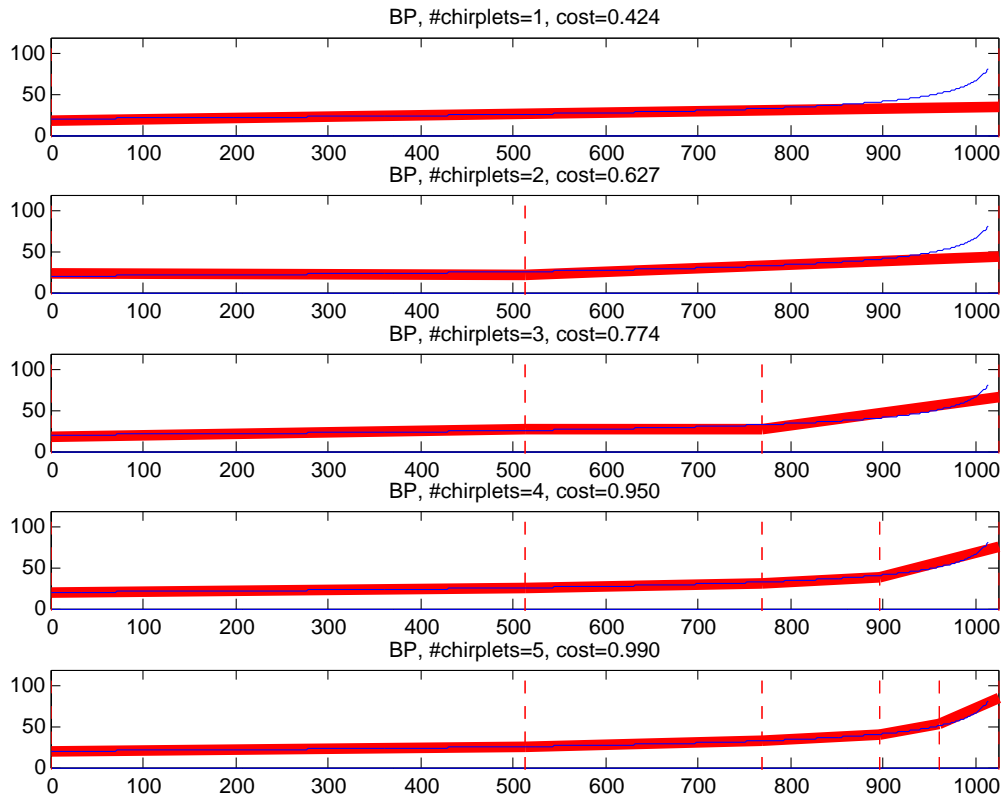}
}
\caption{\small Chirplet paths returned by the BP test for path sizes
  equal to 1, 2, 3, 4, and 5 (the chirplets are adapted to have an
  amplitude varying linearly with time). The signal is a simulated
  gravitational wave.  The cost here is simply the correlation between the
  waveform and the best chirplet path so that a value of 1 indicates a
  perfect match. The horizontal and vertical axes indicate time and
  frequency. The thin line is the `true' instantaneous frequency of the
  waveform. The thick line is the value of the instantaneous frequency
  along the path.}
\label{fig:gwsignal_paths}
\end{figure}

As seen in Figure \ref{fig:gwsignal_paths}, the correlation with the
noiseless waveform is equal to .95 with just 4 chirplets (with linear
time-varying amplitudes) and .99 with just 5 chirplets. So we would
not gain much (if anything at all) by computing inner products with
exponentially many waveforms. 
Another interesting aspect is that the best chirplet path
automatically adapts to the unknown local complexity of the signal; it
uses short templates whenever required and longer templates when the
signal exhibits some coherence over longer periods of time. Here, the
path is refined where the instantaneous frequency starts to rise,
which occurs near the end of the period under study.

\section{Discussion}
\label{sec:discussion}

We have presented a novel and flexible methodology for detecting
nonstationary oscillatory signals. The approach chains together
empirical correlations as to form meaningful signals which may exhibit
very large correlations with the unknown signal we wish to detect.
Our experiments show that our algorithms are very sensitive over very
broad classes of signals. In this section, we discuss further
directions and connections with the work of others.

\subsection{Connection with other works}

While working on this project \cite{IPAM} and writing this paper, we
became aware of the recent and independent work of Chassande-Mottin
and Pai which is similar in spirit to ours \cite{ChassandePai}. In
this paper, the authors also search for a chirplet chain in a graph.
Despite this similitude, our approach is distinct in several
aspects. First, whereas Chassande-Mottin and Pai use chirplets at a
single scale, we use a multiscale dictionary which provides high
flexibility and adaptivity to the unknown structure of the signal (see
Section \ref{sec:why}); the last example in Section
\ref{sec:simulations} also clearly demonstrates the promise of the
multiscale approach for the practical detection of gravitational
waves. Consequently our detection strategy based on the multiple
comparison between test statistics with varying complexities is of
course very different. Second, while we find the best path by dynamic
programming, the best chirplet chain in \cite{ChassandePai} is not the
solution to a tractable optimization problem since the statistic which
needs to be maximized over a set of chirplet paths is not
additive. Therefore the authors need to resort to a series of
approximations involving time-frequency distributions such as the WVD
to obtain an approximate solution.  This makes our approach also
different and more general since the methodology proposed in this
paper may be applied in setups which have nothing to do with chirplets
and chirp detection.

Finally, the aforementioned reference does not address the problem of
detecting chirps with a time varying amplitude, and also assumes that
the noise in the data is white or has been `whitened' in some fashion
(the detection method in \cite{ChassandePai} requires white noise).
In contrast, the statistics in this paper have a natural
interpretation in terms of likelihood ratios, and can be adapted
effortlessly to more sophisticated setups in which the noise may be
colored and in which the amplitude may also be rapidly varying and so
on.  Only the chirplet costs need to be changed while other algorithms
remain the same.

\subsection{Future directions}

It would be of interest to develop a statistical theory of chirp
detection and study whether or not our ideas are provably optimal in a
decision theoretic sense. This would require the development of a
meaningful model of chirps and show that for elements taken from this
model, our methodology nearly obeys the same asymptotic performance
than the minimax test or the Bayes test, should we take the Bayesian
point of view to model our prior knowledge about chirps. We have
investigated these issues and made progress on these problems. The
tools required here seem rather different than those traditionally used
in the mathematical theory of nonparametric detection
\cite{Ingster:NonparamDetectionBook}. We postpone our findings to a
separate report.

There are several extensions to this work which seem worth
investigating. First, we assumed implicitly that the chirp is present
at all times. A more realistic model would need to include the
possibility of chirps with a beginning and an end; i.e. we could only
hear a chirp during a time interval which is shorter than that during
which data is acquired. A more general strategy would then attempt to
detect stretches of data which are not likely to be just noise.  To
isolate such candidate intervals, the dyadic thinking ideas of Castro
et.~al.~\cite{Arias:GeometricObjectsDetection} which consist of
finding promising dyadic intervals and extending these promising
intervals seem especially well suited. Another important extension
would consider detection problems in which not one but several chirps
may be present at once, i.e. the time frequency portrait of the data
may include more than one spectral line.

Our main motivation for this work is the detection of gravitational
waves. While this paper developed a new methodology for this problem,
we did not go as far as testing these ideas on real data.
We are now working on problems posed by real data
(e.g. power line noise, transient events, nonstationarity etc.)
and plan to report on our progress in a future publication.
Of special interest is how one should subsample the
chirplet transform and set the connectivities in the graph to build
sensitive detectors which can deal with large scale data streams while
still demanding manageable computing resources.

\subsection{ChirpLab}
\label{sec:chirplab}

The software package ChirpLab implements the algorithms proposed in
this paper, and is available at {\tt
  http://www.acm.caltech.edu/$\sim$emmanuel/chirplet.html}. It
contains a Matlab implementation of the chirplet transforms and of all
the optimization problems. Several Matlab scripts and a tutorial are
provided to demonstrate how to use this software.

\section{Appendix}
\label{sec:appendix}

\subsection{Real-valued signals}
\label{sec:real}

In our simulations, we considered the detection of complex-valued
chirps and we now rapidly discuss ways to extend the methodology to
real valued data where the signal is of the form $S(t) = A(t)
\cos(\lambda \varphi(t))$ with unknown phase and amplitude. Again, the
idea is to build a family of real-valued chirplets which exhibit good
local correlations with the signal. To do this, we could consider
chirplets with quadratic phase $a_\mu t^2/2 + b_\mu t + c_\mu$ and
build a graph in which connectivities impose regularity assumptions on
the phase function. The downside with this approach is that for each
chirplet, one would need to introduce the extra phase-shift parameter
$c_\mu$, which would increase the size of the dictionary and of the
graph. This is not desirable.

A much better strategy is as follows: we parameterize chirplets in the
same way with $v = (I, \mu)$ where $I$ is the time support of a
chirplet and $a_\mu t + b_\mu$ is the instantaneous frequency, and
define the chirplet cost by
\begin{equation}
  \label{eq:real-cost}
  C(v) := \max_{c} \frac{\left|\sum_{t \in I} y_t \cos(a_\mu t^2/2 +
    b_\mu t + c)\right|^2}{\sum_{t \in I} \cos^2(a_\mu t^2/2 + b_\mu t
  + c)}.
\end{equation}
That is, we simply select the phase shift which maximizes the
correlation (note that with complex data, the corresponding ratio
$|\sum y_t \exp(\i (a_\mu t^2/2 + b_\mu t + c))|^2/\sum |\exp(\i(a_\mu
t^2/2 + b_\mu t + c))|^2$ is, of course, independent of $c$).  One can
use simple trigonometric identities and write the numerator and
denominator in \eqref{eq:real-cost} as
\[
A^2 \cos^2 c - 2 AB \sin c \cos c + B^2 \sin^2 c, \quad C^2 \cos^2 c -
2 D \sin c \cos c + E^2 \sin^2 c, 
\]
where with  $\vf_\mu(t) = a_\mu t^2/2 + b_\mu t$, 
\[
A + \i B = \sum y_t \, e^{\i \vf_\mu(t)},\\
\]
and
\[
C^2 = \sum \cos^2 \vf_\mu(t), \quad D = \sum \cos \vf_\mu(t) \sin
\vf_\mu(t), \quad E^2 = \sum \sin^2 \vf_\mu(t).
\]
Note that $A + \i B$ is nothing else than the chirplet coefficient of
the data and $C, D$, and $E$ can be computed off-line.  There is an
analytic formula for finding the value of $\cos c$ (or $\sin c$) that
maximizes the ratio as a function of $A, B, C, D$, and $E$. This
extends to the more sophisticated setups discussed in Section
\ref{sec:extensions}.

Finally, there are further approximations which one could use as
well. Observe the expansion of the denominator in \eqref{eq:real-cost}
\[
\sum_{t \in I} \cos^2 (\varphi_\mu(t) + c) = |I|/2 + 1/2 \, \sum_{t \in
  I} \cos (2\varphi_\mu(t) + 2c),
\]
where $|I|$ is here the number of time samples in $I$.  Then for most
chirplets (when the support contains a large number of oscillations),
the second term in the right-hand side is negligible compared to the
first. Assuming that the denominator is about equal to $|I|/2$ for all
phase shifts $c$, we would then simply maximize the numerator in
\eqref{eq:real-cost}. A simple calculation shows that $e^{i c} = (A +
\i B)/\sqrt{A^2 + B^2}$ and
\[
C(v) \approx 2 \, \left| \sum_{t \in I} y_t e^{-\i (a_\mu t^2 + b_\mu
    t)}/\sqrt{|I|} \right|^2.
\]
(the ``$\approx$'' symbol indicates the approximation in the
denominator). Hence, the real-valued cost is just about twice the
usual complex-valued cost.

\subsection{MCCTR algorithms}
\label{sec:mcttr}

In this section, we briefly argue that one can compute the MCTTR
introduced in Section \ref{sec:variations} efficiently
\cite{Ahuja:NetworkFlows,Dantzig:MCTR}. Assume that $p^\star$ is the
maximum value of $\sum_{i \in W} c(i,j)/|W|$ (with optimal solution
$W^\star$) and that we have a lower bound $p_0$ on $p^\star$ (a
trivial lower bound for the chirplet problem is $p_0 = 0$). Suppose
that $W_0$ solves the SP problem with modified costs $c_0(i,j) =
c(i,j) - p_0$. Then there are three possible cases, and we will rule
one out:
\begin{enumerate}[i)]
\item $\sum_{W_0} c_0(i,j) < 0$. Then $\sum_{W} c_0(i,j) \leq
  \sum_{W_0} c_0(i,j) < 0$ for all paths $W$ and $\sum_{W^\star}
  c(i,j)/|W^*| < p_0 \le p^\star$. This is a contradiction and this
  case never comes up.

\item $\sum_{W_0} c_0(i,j) = 0$. Then $\sum_{W} c_0(i,j)
  \leq \sum_{W_0} c_0(i,j) = 0$ and, hence, $\sum_{W} c(i,j)/|W|
  \leq p_0$ for all paths $W$. We conclude that $p_0 = p^\star$.

\item $\sum_{W_0} c_0(i,j) > 0$. Then $\sum_{W_0} c(i,j)/|W_0| > p_0$
  and we have a tighter lower bound on $p^*$. Take $p_1 = \sum_{W_0}
  c(i,j)/|W_0|$ and repeat with the new costs $c_1(i,j) = c(i,j) -
  p_1$.
\end{enumerate}
The MCTTR algorithm solves a sequence of SP problems, and visits a
subset of the vertices on the boundary of the convex hull of the
points $(|W|, C(W))$ until it finds the optimal trade-off. The number
of vertices is of course bounded by the maximum possible length
$\ell_{\max}$ of the path. In practice, the MCTTR converges after just
a few iterations---between 4 and 6 in our simulations.

\bibliographystyle{plain}
\bibliography{ChirpDetectionBib}

\begin{thebibliography}{10}

\bibitem{LIGO}
A.~{Abramovici}, W.~E. {Althouse}, R.~W.~P. {Drever}, Y.~{Gursel},
  S.~{Kawamura}, F.~J. {Raab}, D.~{Shoemaker}, L.~{Sievers}, R.~E. {Spero}, and
  K.~S. {Thorne}.
\newblock {LIGO - The Laser Interferometer Gravitational-Wave Observatory}.
\newblock {\em Science}, 256:325--333, April 1992.

\bibitem{Ahuja:NetworkFlows}
R.~K Ahuja, T.~L. Magnanti, and J.~B. Orlin.
\newblock {\em Network flows. Theory, algorithms and applications}.
\newblock Prentice-Hall, New-York, 1993.

\bibitem{allen}
B.~Allen.
\newblock Gravitational radiation analysis and simulation package {(GRASP)}.
\newblock Technical report, Department of Physics, University of Wisconsin,
  Milwaukee, P.O. Box 413, Milwaukee, WI 53201, 2000.
\newblock Available at {\tt
  http://www.lsc-group.phys.uwm.edu/$\sim$ballen/grasp-distribution/index.html%
}.

\bibitem{anderson:tfgwdetection}
W.~G. Anderson and R.~Balasubramanian.
\newblock Time-frequency detection of gravitational waves.
\newblock {\em Physical Review D (Particles, Fields, Gravitation, and
  Cosmology)}, 60(10):102001, 1999.

\bibitem{DetectPath}
E.~Arias-Castro, E.~J. Cand\`es, and H.~Helgason.
\newblock Several examples of near optimal path detection from noisy data.
\newblock 2006.

\bibitem{Arias:GeometricObjectsDetection}
E.~Arias-Castro, D.~L. Donoho, and X.~Huo.
\newblock Near-optimal detection of geometric objects by fast multiscale
  methods.
\newblock {\em IEEE Trans. Inform. Theory}, 51(7):2402--2425, 2005.

\bibitem{Arias:FilamentDetection}
E.~Arias-Castro, D.~L. Donoho, and X.~Huo.
\newblock Adaptive multiscale detection of filamentary structures in a
  background of uniform random points.
\newblock {\em Ann. Statist.}, 34(1), 2006.

\bibitem{BaraniukJones}
R.~G. Baraniuk and D.~L. Jones.
\newblock Shear madness: New orthonormal bases and frames using chirp
  functions.
\newblock {\em IEEE Transactions on Signal Processing}, 41(12):3543--3548,
  1993.

\bibitem{IPAM}
E.~J Cand\`es, P.~Charlton, and H.~Helgason.
\newblock Chirplets: recovery and detection of chirps.
\newblock 2004.
\newblock Presentation at the Institute of Pure and Applied Mathematics.

\bibitem{GWs}
E.~J Cand\`es, P.~Charlton, and H.~Helgason.
\newblock Detecting gravitational waves via chirplet path pursuit.
\newblock 2006.

\bibitem{ChassandreFlandrin}
E.~Chassande-Mottin and P.~Flandrin.
\newblock On the time-frequency detection of chirps.
\newblock {\em Appl. Comput. Harmon. Anal.}, 6(2):252--281, 1999.

\bibitem{ChassandePai}
E.~Chassande-Mottin and A.~Pai.
\newblock Best chirplet chain: near-optimal detection of gravitational wave
  chirps.
\newblock {\em Phys. Rev. D}, 73(4):042003 --- 1--25, 2006.

\bibitem{Dantzig:MCTR}
G.~B. Dantzig, W.~O. Blattner, and M.~R. Rao.
\newblock Finding a cycle in a graph with minimum cost to time ratio with
  application to a ship routing problem.
\newblock In {\em Theory of Graphs (Internat. Sympos., Rome, 1966)}, pages
  77--83. Gordon and Breach, New York, 1967.

\bibitem{Donoho:Beamlets}
D.~L. Donoho and X.~Huo.
\newblock Beamlets and multiscale image analysis.
\newblock In {\em Multiscale and multiresolution methods}, volume~20 of {\em
  Lect. Notes Comput. Sci. Eng.}, pages 149--196. Springer, Berlin, 2002.

\bibitem{Whistlers1}
R.~A. Helliwell.
\newblock {\em Whistlers and Related Atmospheric Phenomena}.
\newblock Stanford University Press, 1965.

\bibitem{Ingster:NonparamDetectionBook}
Y.~I. Ingster and I.~A. Suslina.
\newblock {\em Nonparametric goodness-of-fit testing under {G}aussian models},
  volume 169 of {\em Lecture Notes in Statistics}.
\newblock Springer-Verlag, New York, 2003.

\bibitem{Joksch:CSP}
H.~C. Joksch.
\newblock The shortest route problem with constraints.
\newblock {\em J. Math. Anal. Appl.}, 14:191--197, 1966.

\bibitem{MannHaykin}
S.~Mann and S.~Haykin.
\newblock The chirplet transform: Physical considerations.
\newblock {\em IEEE Transactions on Signal Processing}, 43(11):2745--2761,
  1995.

\bibitem{MeyerLewis}
Y.~Meyer.
\newblock {\em Oscillating patterns in image processing and nonlinear evolution
  equations}, volume~22 of {\em University Lecture Series}.
\newblock American Mathematical Society, Providence, RI, 2001.
\newblock The fifteenth Dean Jacqueline B. Lewis memorial lectures.

\bibitem{Torresani:ChirpDetection}
M.~Morvidone and B.~Torresani.
\newblock Time scale approach for chirp detection.
\newblock {\em Int. J. Wavelets Multiresolut. Inf. Process.}, 1(1):19--49,
  2003.

\bibitem{owen:matchedfiltering}
B.~J. Owen and B.~S. Sathyaprakash.
\newblock Matched filtering of gravitational waves from inspiraling compact
  binaries: Computational cost and template placement.
\newblock {\em Physical Review D (Particles, Fields, Gravitation, and
  Cosmology)}, 60(2):022002, 1999.

\bibitem{Whales}
J.~E. Reynolds~III and S.~A. Rommel.
\newblock {\em Biology of Marine Mammals}.
\newblock Smithsonian Institution Press, 1999.

\bibitem{Ullman}
A.~Sha'~ashua and S.~Ullman.
\newblock The detection of globally salient structures using a locally
  connected network.
\newblock In {\em Proceedings of the Second International Conference on
  Computer Vision}, pages 321--327, 1988.

\bibitem{Bats2}
J.~Simmons.
\newblock Echolocation in bats: signal processing of echoes for target range.
\newblock {\em Science}, 171(974):925--928, 1971.

\bibitem{Bats1}
J.~Simmons.
\newblock The resolution of target range by echolocating bats.
\newblock {\em The Journal of the Acoustical Society of America}, 54:157--173,
  1973.

\bibitem{thorne:300yearsofgravitation}
K.~S. Thorne.
\newblock Graviational radiation.
\newblock In Stephen~W. Hawking and Werner Israel, editors, {\em 300 Years of
  Gravitation}, pages 330--358. Cambridge University Press, Cambridge, England,
  1987.

\bibitem{MCP}
P.~Westfall and S.~Young.
\newblock {\em Resampling-based Multiple Testing: Examples and Methods for
  P-value Adjustment}.
\newblock Wiley, New York, 1993.

\end{thebibliography}

\end{document}